\documentclass[a4paper,final,5p,times,twocolumn]{elsarticle}
\usepackage{amsmath}
\usepackage{amsfonts}
\usepackage{amssymb}
\usepackage{color}
\usepackage{fixltx2e}
\usepackage{hyperref}
\usepackage{amssymb}
\usepackage{amsthm}
\usepackage{longtable}
\usepackage{graphicx}
\usepackage[dvips]{epsfig}
\usepackage{bbm}
\usepackage{dsfont}
\usepackage{flushend}
\usepackage{textcomp}
\usepackage{hyperref}
\usepackage{doi}

\biboptions{sort&compress}

\newtheorem{theorem}{Theorem}[section]
\newtheorem{lemma}[theorem]{Lemma}
\newtheorem{proposition}[theorem]{Proposition}

\newtheorem{example}[theorem]{Example}

\newtheorem{remark}[theorem]{Remark}

\newcommand{\carre}{\hfill $\blacksquare$}
\newcommand{\carrew}{\hfill $\square$}

\makeatletter
\newcommand{\labitem}[2]{%
\def\@itemlabel{\text{#1}}
\item
\def\@currentlabel{#1}\label{#2}}
\makeatother

\journal{Automatica}

\begin{document}

\begin{frontmatter}



\title{\vspace{-1cm}Observers for Invariant Systems on Lie Groups with Biased Input Measurements and Homogeneous Outputs\tnoteref{labelacknowledge}}
\tnotetext[labelacknowledge]{This work was partially supported by the Australian Research Council through the ARC Discovery Project DP120100316 "Geometric Observer Theory for Mechanical Control Systems".\\
This paper was published in Automatica \cite{khosravian2015observers}, \doi{10.1016/j.automatica.2015.02.030}\\
\textcopyright 2015, Elsevier.  This manuscript version is made available under the Creative Commons Attribution-NonCommercial-NoDerivatives CC-BY-NC-ND 4.0 license \url{http://creativecommons.org/licenses/by-nc-nd/4.0/}}

\author[label1]{Alireza Khosravian}
\author[label1]{Jochen Trumpf}
\author[label1]{Robert Mahony}
\author[label2]{Christian Lageman}
\address[label1]{Research School of Engineering, Australian National University, Canberra ACT 2601, Australia. (e-mails:  alireza.khosravian@anu.edu.au; jochen.trumpf@anu.edu.au; robert.mahony@anu.edu.au)}
\address[label2]{Mathematical Institute, University of W\"{u}rzburg, 97074 W\"{u}rzburg, Germany. (e-mail: christian.lageman@mathematik.uni-wuerzburg.de)\vspace{-5mm}}

%\vspace{-1cm}
\begin{abstract}
This paper provides a new observer design methodology for invariant systems whose state evolves on a
Lie group with outputs
in a collection of related homogeneous spaces and where the measurement of system input is corrupted by an unknown
constant bias.
The key contribution of
the paper is to study the combined state and input bias estimation problem in the general setting of Lie groups,
a question for which only case studies of specific Lie groups are currently available.
We show that any candidate observer (with the same state space dimension as the observed system) results in
non-autonomous error dynamics, except in the trivial case where the Lie-group is Abelian.
This precludes the application of the standard non-linear observer design methodologies available in the
literature and leads us to propose a new design methodology based on employing invariant cost functions
and general gain mappings. We provide a rigorous and general stability analysis for the case where the underlying Lie group allows a faithful matrix representation.
We demonstrate our theory in the example of rigid body pose estimation and show that the
proposed approach unifies two competing pose observers published in prior literature.
\end{abstract}


\end{frontmatter}

% \linenumbers

\section{Introduction}
The study of dynamical systems on Lie groups has been an active research area for the past decade.
Work in this area is motivated by applications in analytical mechanics, robotics and geometric control for mechanical systems \cite{jurdjevic1997geometric,bloch2003nonholonomic,agrachev2004control,bullo2005geometric}. Many mechanical systems carry a natural symmetry or invariance structure expressed as invariance properties of their dynamical models under transformation by a symmetry group. 
For totally symmetric kinematic systems, the system can be lifted to an invariant system on the symmetry group \cite{mahony2013observers}. 
In most practical situations, obtaining a reliable measurement of the internal states of such physical systems directly is not possible and it is necessary to use a state observer. 

Systematic observer design methodologies for invariant systems on Lie groups have been proposed that lead to strong stability and robustness properties. Specifically, Bonnabel \emph{et al.} \cite{bonnabel2008symmetry,bonnabel2009non,bonnabel2008non} consider observers which consist of a copy of the system and a correction term, along with a constructive method to find suitable symmetry-preserving correction terms.
The construction utilizes the invariance of the system and the moving frame method,
leading to local convergence properties of the observers. The authors propose methods in \cite{lageman2010gradient,Lageman09,lageman2008observer} to achieve almost globally convergent observers. A key aspect of the design approach proposed in \cite{lageman2010gradient,Lageman09,lageman2008observer} is the use of the invariance properties of the system to ensure that the error dynamics are globally
defined and are autonomous. This leads to a straight forward stability analysis and excellent performance in practice. More recent extensions to early work in this area was the consideration of output measurements where a partial state measurement is generated by an action of the Lie group on a homogeneous output space \cite{mahony2013observers,lageman2008observer,Lageman09,bonnabel2008symmetry,bonnabel2009non,bonnabel2008non,khosravian2013bias}.
Design methodologies exploiting symmetries and invariance of the system can be applied to many real world scenarios such as attitude estimator design on the Lie group $\textsf{SO}(3)$ \cite{Mahony08,grip2012attitude,bras2011nonlinear,Vasconcelos08a,Khosravian10,bonnabel2009non}, pose estimation on the Lie group $\textsf{SE}(3)$ \cite{baldwin2007complementary,vasconcelos2010nonlinear,hua2011observer,rehbinder2003pose}, homography estimation on the Lie group $\textsf{SL}(3)$ \cite{hamel2011homography}, and motion estimation of chained systems on nilpotent Lie groups \cite{leonard1995motion} (e.g. front-wheel drive cars or kinematic cars with $k$ trailers).

All asymptotically stable observer designs for kinematic systems on Lie groups depend on a measurement of system input.
In practice, measurements of system input are often corrupted by an unknown bias that must be estimated and compensated to achieve good observer error performance.
The specific cases of attitude estimation on $\textsf{SO}(3)$ and
pose estimation on $\textsf{SE}(3)$ have been studied independently,
and methods have been proposed for the concurrent estimation of state and input measurement bias
\cite{Mahony08,Vasconcelos08a,vasconcelos2010nonlinear}. These methods strongly depend on particular
properties of the specific Lie groups $\textsf{SO}(3)$ or $\textsf{SE}(3)$ and do not directly generalize to general Lie groups. 
To the authors' knowledge, there is no existing work on combined state and input bias estimation for general classes of invariant systems.

In this paper, we tackle the problem of observer design for general invariant systems on Lie groups with
homogeneous outputs when the measurement of system input is corrupted by an unknown constant bias.
The observer is required to be implementable based on available sensor measurements; the system input in the
Lie algebra, corrupted by an unknown bias, along with a collection of
partial state measurements (i.e. outputs) that ensure observability of the state. 
For bias free input measurements, it is always possible to obtain autonomous dynamics for the standard error \cite{lageman2010gradient,Lageman09,lageman2008observer}, and previous observer design methodologies for systems on Lie groups rely on
the autonomy of the resulting error dynamics. 
However, for concurrent state and input measurement bias estimation, we show that \emph{any} implementable candidate observer (with the same state space dimension as the observed system) yields non-autonomous error dynamics unless the Lie group is Abelian (Theorem \ref{theo_autonomy}).
This result explains why the previous general observer design methodologies for the bias-free case do not apply and why the special cases considered in prior works \cite{hua2011observer,vasconcelos2010nonlinear} do not naturally lead to a general theory.

We go on to show that, despite the nonlinear and non-autonomous nature of the error dynamics, there is a natural choice of observer for which we can prove exponential stability of the error dynamics (Theorems \ref{theo_observer_stability} and \ref{theo_observer_exp_stability}).
The approach taken employs a general gain mapping applied to the differential of a cost function rather than the more restrictive gradient-like innovations used in prior work  
\cite{lageman2010gradient,Lageman09,lageman2008observer,khosravian2013bias}. We also propose a systematic method for construction of invariant cost functions based on lifting costs defined on the homogeneous output spaces (Proposition \ref{lemma_cost}).
To demonstrate the generality of the proposed approach we consider the problem of rigid body pose estimation using landmark measurements when the measurements of linear and angular velocity are corrupted by constant unknown biases.
We show that for specific choices of gain mappings the resulting observer specializes to either the gradient-like observer of \cite{hua2011observer} or the non-gradient pose estimator proposed in \cite{vasconcelos2010nonlinear}, unifying these two  state-of-the-art application papers in a single framework that applies to any invariant kinematic system on a Lie-group. Stability of estimation error is proved for the case where the Lie group allows a faithful matrix representation.

The paper is organized as follows. 
After briefly clarifying our notation in Section \ref{preliminaries},
we formulate the problem in Section \ref{sec_problem}. A standard estimation error is defined and autonomy
of the resulting error dynamics is investigated in section \ref{sec_autonomy}. We introduce the
proposed observer in Section \ref{sec_observer_design} and investigate the stability of observer error dynamics.
Section \ref{sec_realizability_grad} is devoted to the systematic construction of invariant cost functions.
A detailed example in Section \ref{sec_SE3_exmple} and brief conclusions in Section \ref{sec_conclusion}
complete the paper. A preliminary version of this work was presented at the CDC 2013 \cite{khosravian2013bias}. This manuscript was published in Automatica \cite{khosravian2015observers}. In addition to the material presented in \cite{khosravian2015observers}, this paper contains detailed proof of theorems as well as detailed mathematical derivations of application examples.

\section{Notations and Definitions}  \label{preliminaries}
Let $G$ be a finite-dimensional real connected Lie group with associated Lie algebra $\mathfrak{g}$. Denote the identity element of $G$ by $I$. Left (resp. right) multiplication of $X\in G$ by $S\in G$ is denoted by $L_S X=S X$ (resp. $R_S X=X S$). The Lie algebra $\mathfrak{g}$ can be identified with the tangent space at the identity element of the Lie group, i.e. $\mathfrak{g} \cong T_I G$. For any $u \in \mathfrak{g}$, one can obtain a tangent vector at $S \in G$ by left (resp. right) translation of $u$ denoted by $S [u]:=T_I L_S [u] \in T_S G$ (resp. $[u] S:= T_I R_S [u] \in T_S G$). The element inside the brackets $[.]$ denotes the vector on which a linear mapping (here the tangent map $T_I L_S\colon\mathfrak{g} \to T_S G$ or $T_I R_S\colon \mathfrak{g} \to T_S G$) acts. The adjoint map at the point $S\in G$ is denoted by $\text{Ad}_S\colon\mathfrak{g} \to \mathfrak{g}$ and is defined by $\text{Ad}_S [u]:=S[u]S^{-1}=T_S R_{S^{-1}} [ T_I L_S [u]]=T_S R_{S^{-1}} \circ T_I L_S [u]$ where $\circ$ denotes the composition of two maps. For a finite-dimensional vector space $V$, we denote its corresponding dual and bidual vector spaces by $V^*$ and $V^{**}$ respectively. A linear map $F\colon V^* \to V$ is called positive definite if $v^* [F [v^*]] > 0$ for all $0 \ne v^* \in V^*$. The dual of $F$ is denoted by $F^*\colon V^* \to V^{**}$ and is defined by $F^*[v^*] =v^* \circ F$. The linear map $F$ is called symmetric (resp. anti-symmetric) if $v^*[F[w^*]]=w^*[F[v^*]]$ (resp. $v^*[F[w^*]]=-w^*[F[v^*]]$) for all $v^*,w^* \in V^*$, and it is called symmetric positive definite if it is symmetric and positive definite. We can extend the above notion of symmetry and positiveness to linear maps $H\colon W \to W^*$ as well. Defining $V:=W^*$, $H$ is called positive definite if $H^*\colon V^* \to V$ is positive definite and it is called symmetric if $H^*$ is symmetric. Positive definite cost functions on manifolds are also used in the paper and should not be mistaken with positive definite linear maps.

\section{Problem Formulation} \label{sec_problem}
We consider a class of left invariant systems on $G$ given by
\begin{equation} \label{group_left_dyn}
\dot X(t) = X(t) u(t),~~~~~~X(t_0)=X_0,
\end{equation}
where $u\in \mathfrak{g}$ is the system input and $X\in G$ is the state. Although the ideas presented in this paper are based on the above left invariant dynamics, they can easily be modified for right invariant systems as was done for instance in \cite{lageman2010gradient}. We assume that $u\colon\mathds{R}^+\to \mathfrak{g}$ is continuous and hence a unique solution for (\ref{group_left_dyn}) exists for all $t\ge t_0$ \cite{jurdjevic1972control}. In most kinematic mechanical systems, $u$ models the velocity of physical objects. Hence, it is reasonable to assume that $u$ is bounded and continuous.

Let $M_i,~i=1,\ldots,n$ denote a collection of $n$ homogeneous spaces of $G$, termed \emph{output spaces}. Denote the outputs of system (\ref{group_left_dyn}) by $y_i \in M_i$. Suppose each output provides a partial measurement of $X$ via
\begin{align} \label{outputs}
y_i=h_i(X,\mathring{y}_i)
\end{align}
where $\mathring{y}_i \in M_i$ is the constant (with respect to time) reference output associated with $y_i$ and $h_i$ is a right action of $G$ on $M_i$, i.e. $h_i(I, y_i)=y_i$ and $h_i(X S, y_i)=h_i(S,h_i(X,y_i))$ for all $y_i \in M_i$ and all $X,S\in G$. To simplify the notation, we define the combined output $y:=(y_1,\ldots,y_n)$, the combined reference output $\mathring y:=(\mathring{y}_1,\ldots,\mathring{y}_n)$, and the combined right action $h(X,\mathring y):=(h_1(X,\mathring{y}_1),\ldots,h_n(X,\mathring{y}_n))$. The combined output $y$ belongs to the orbit of $G$ acting on the product space $M_1 \times M_2 \times \ldots \times M_n$ containing $\mathring y$, that is $M:=\{y \in M_1 \times M_2 \times \ldots \times M_n |~ y=h(X,\mathring y),~X\in G \} \subset M_1 \times M_2 \times \ldots \times M_n$. Note that the combined action $h$ of $G$ defined above is transitive on $M$. Hence, $M$ is a homogeneous space of $G$ while $M_1 \times M_2 \times \ldots \times M_n$ is not necessarily a homogeneous space of $G$ \cite{kobayashi1996foundations}.

We assume that measurements of the system input are corrupted by a constant unknown additive bias. That is
\begin{align} \label{u_y}
u_y&=u+b
\end{align}
where $u_y\in \mathfrak{g}$ is the measurement of $u$ and $b\in \mathfrak{g}$ is the unknown bias. In practice, bias is slowly time-varying but it is common to assume that $b$ is constant for observer design and analysis.

We investigate the observer design problem for concurrent estimation of $X$ and $b$.
The observer should be \emph{implementable} based on sensor measurements.
This is important since the actual state $X \in G$ and the actual input $u \in \mathfrak{g}$ are not
available for measurement and only the partial measurements $y_1,\ldots,y_n$ and the biased input $u_y$
are directly measured. We consider the following general class of implementable observers
with the same state space dimension as the observed system.
\begin{subequations} \label{observer_implementable}
\begin{align}
\label{observer_implementable_group} \dot {\hat X} &=\gamma(\hat X, y, \mathring y, \hat b,u_y,t)\\
\label{observer_implementable_bias} \dot {\hat b} &=\beta(\hat X, y,\mathring y,\hat b,u_y,t)
\end{align}
\end{subequations}
where $\hat X$ and $\hat b$ are the estimates of $X$ and $b$, respectively,
and $\gamma\colon G \times M \times M \times  \mathfrak{g} \times \mathfrak{g} \times \mathds{R} \to T G$
and $\beta\colon G \times  M \times M \times \mathfrak{g} \times \mathfrak{g} \times \mathds{R} \to \mathfrak{g}$
are parameterized vector fields on $G$ and $\mathfrak{g}$, respectively. Note that $\hat X, y, \mathring y, \hat b,u_y$ and $t$ are all available for implementation of the observer in practical scenarios. We refer to (\ref{observer_implementable_group}) and (\ref{observer_implementable_bias}) as the \emph{group estimator} and the \emph{bias estimator}, respectively.

\begin{example}[Pose Estimation on $\textsf{SE}(3)$] \label{example_SE3}
Estimating the position and attitude of a rigid body has been investigated by a range of authors during the past decades, see, e.g., \cite{baldwin2007complementary,hua2011observer,rehbinder2003pose,vasconcelos2010nonlinear,vik2001nonlinear}. The full 6-DOF pose kinematics of a rigid body can be modeled as an invariant system on the special Euclidean group  $\textsf{SE}(3)$ \cite{rehbinder2003pose,vasconcelos2010nonlinear,vik2001nonlinear,lageman2010gradient}. The Lie group $\textsf{SE}(3)$ has a representation as a semi-direct product of $\textsf{SO}(3)$ and $\mathds{R}^3$ given by $\textsf{SE}(3)=\{ (R,p)|~R \in \textsf{SO}(3),~ p \in \mathds{R}^3 \}$ \cite{bourbaki1989lie}. Consider the group multiplication on $\textsf{SE}(3)$ given by $R_{(S,q)} (R,p)=L_{(R,p)} (S,q)=(RS,p+Rq)$ for any $(R,p),(S,q) \in \textsf{SE}(3)$. The identity element of $\textsf{SE}(3)$ is represented by $(I_{3\times 3},0_3)$ and the inverse of an element $(R,p)\in \textsf{SE}(3)$ is given by $(R,p)^{-1}=(R^\top,-R^\top p)$. The Lie algebra of $\textsf{SE}(3)$ is identified with $\mathfrak{se}(3)=\{(\Omega,V)|~\Omega \in \mathfrak{so}(3),~V\in \mathds{R}^3\}$ where $\mathfrak{so}(3)$ denotes the Lie algebra of $\textsf{SO}(3)$ represented as the set of skew-symmetric $3\times 3$ matrices with zero trace.

Let $R$ be a rotation matrix corresponding to the rotation from the body-fixed frame to the inertial frame and suppose that $p$ represents the position of the rigid body with respect to the inertial frame and expressed in the inertial frame. The left-invariant kinematics of a rigid body on $\textsf{SE}(3)$ is formulated as
\begin{align} \label{SE3_kinematics}
(\dot R, \dot p)=T_I L_{(R,p)} (\Omega,V)=(R\Omega, RV)
\end{align}
where $\Omega$ resp. $V$ represent the angular velocity resp. linear velocity of the rigid body with respect to the inertial frame expressed in the body-fixed frame. Here, the group element is $X=(R,p)\in \textsf{SE}(3)$ and the system input is $u=(\Omega,V) \in \mathfrak{se}(3)$. Denote the measurement of the system input by $(\Omega_y,V_y)\in \mathfrak{se}(3)$ and assume that it is corrupted  by an unknown constant bias $(b_\omega,b_v)\in \mathfrak{se}(3)$ such that $(\Omega_y,V_y)=(\Omega+b_\omega,V+b_v)$. Suppose that positions of $n$ points with respect to the body-fixed frame are measured by on-board sensors and denote these measurements by $y_1,\ldots,y_n \in \mathds{R}^3$. Denote the positions of these points with respect to the inertial frame by $\mathring y_i,~i=1,\ldots,n \in \mathds{R}^3$ and assume these positions are known and constant. The output model for such a set of measurements is given by
\begin{align} \label{SE3_measurement}
y_i=h_i((R,p),\mathring{y}_1)=R^\top \mathring y_i - R^\top p,~~~i=1,\ldots,n
\end{align}
where $h_i$ is a right action of $\textsf{SE}(3)$ on the homogeneous output space $M_i:=\mathds{R}^3$. A practical example of measurements modeled by (\ref{SE3_measurement}) is vision based landmark readings where the landmarks are fixed in the inertial frame, leading to constant $\mathring y_i,~i=1,\ldots,n$ \cite{vasconcelos2010nonlinear}. The pose estimation problem is to estimate $R$ and $p$ together with the input biases $b_\omega$ and $b_v$. \carrew
\end{example}

\section{Error Definition and Autonomy of Error Dynamics} \label{sec_autonomy}
We consider the following right-invariant group error,
\begin{align}  \label{right_error}
E=\hat X X^{-1} \in G,
\end{align}
as was proposed in \cite{lageman2010gradient,mahony2013observers}. The above error resembles the usual error $\hat x-x$ used in classical observer theory when $\hat x , x$ belong to a vector space. We have $\hat X=X$ if and only if $E=I$. We also consider the following bias estimation error
\begin{align} \label{tilde_b}
\tilde b=\hat b-b \in \mathfrak{g}.
\end{align}

We are interested to see when an observer of the general form (\ref{observer_implementable}) produces \textit{autonomous} error dynamics since that would enable straight-forward stability analysis. When the measurement of system input is bias free, implementable observers of the form (\ref{observer_implementable_group}) have been proposed that produce autonomous group error dynamics $\dot E$ \cite{lageman2010gradient}. In this section, we show that when the measurement of system input is corrupted by bias, any implementable observer of the form (\ref{observer_implementable}) produces non-autonomous error dynamics for general Lie groups, and it can only produce autonomous error dynamics for Abelian Lie groups. To prove this result, we note that the observer (\ref{observer_implementable}) can be rewritten into the form
\begin{subequations} \label{observer}
\begin{align}
\label{group_estimator} \dot {\hat X} &=\hat X [u_y -\hat b]-\alpha_{\mathring y} (\hat X, y, \hat b,u_y,t),\\
\label{bias_estimator} \dot {\hat b}&=\beta_{\mathring y}(\hat X, y,\hat b,u_y,t),
\end{align}
\end{subequations}
where $\alpha_{\mathring y}\colon G \times M \times \mathfrak{g} \times \mathfrak{g} \times \mathds{R} \to T G$ and $\beta_{\mathring y}\colon G \times M \times \mathfrak{g} \times \mathfrak{g} \times \mathds{R} \to \mathfrak{g}$ are parameterized vector fields on $G$ and $\mathfrak{g}$, respectively, and $\mathring y$ is now interpreted as a parameter for $\alpha_{\mathring y}$ and $\beta_{\mathring y}$.

\begin{theorem} \label{theo_autonomy}
Consider the observer (\ref{observer}) for the system (\ref{group_left_dyn})-(\ref{u_y}). The error dynamics $(\dot E,\dot{\tilde b})$ is autonomous if and only if all of the following conditions hold;
\begin{enumerate}
\labitem{(a)}{con_autonomy_indep_param} $\alpha_{\mathring y}$ and $\beta_{\mathring y}$ do not depend on $\hat b$, $u_y$ and $t$.
\labitem{(b)}{con_autonomy_alpha_invar} The vector field $\alpha_{\mathring y}$ is right equivariant in the sense that $T_{\hat X} R_Z \alpha_{\mathring y}(\hat X,y)=\alpha_{\mathring y}(\hat X Z,h(Z,y))$ for all $\hat X,Z\in G$ and all $y \in M$.
\labitem{(c)}{con_autonomy_beta_invar}{The vector field $\beta_{\mathring y}$ is right invariant in the sense that $\beta_{\mathring y}(\hat X,y)=\beta_{\mathring y}(\hat X Z,h(Z,y))$ for all $\hat X,Z\in G$ and all $y \in M$.}
\labitem{(d)}{con_autonomy_adj_trivial}{For all $Z \in G$ the adjoint map $\text{Ad}_Z\colon \mathfrak{g}\to \mathfrak{g}$ is the identity map.}\carrew
\end{enumerate}
\end{theorem}

\textit{Proof: } In view of (\ref{group_left_dyn}) and (\ref{observer}), differentiating $E=\hat X X^{-1}$ and $\tilde b=\hat b-b$ with respect to time yields
\begin{subequations} \label{error}
\begin{align}
\dot E &=-T_{\hat X} R_{X^{-1}} \alpha_{\mathring y}(\hat X, y, \hat b,u_y,t)-T_I R_{E} \text{Ad}_{\hat X} \tilde b\\
\dot{\tilde b} &=\beta_{\mathring y}(\hat X, y,\hat b,u_y,t).
\end{align}
\end{subequations}
If the conditions \ref{con_autonomy_indep_param} to \ref{con_autonomy_adj_trivial} of the Theorem hold, the error dynamics will be simplified to
\begin{subequations}
\begin{align}
\dot E &\!=\!- \alpha_{\mathring y}(\hat X X^{-1}\!, h(X^{-1}\!,y))\!-\!T_I R_{E} \tilde b\!=\!- \alpha_{\mathring y}(E, \mathring y)\!-\!T_I R_{E} \tilde b,\\
\dot{\tilde b} &=\beta_{\mathring y}(\hat X X^{-1}, h(X^{-1},y))=\beta_{\mathring y}(E, \mathring y),
\end{align}
\end{subequations}
which are autonomous.

Conversely, assume that the error dynamics (\ref{error}) are autonomous. Then there exist functions $F_{\mathring y}\colon G\times \mathfrak{g} \to TG$ and $H_{\mathring y}\colon G\times \mathfrak{g} \to \mathfrak{g}$ such that for all $X,\hat X \in G$, $y \in M$, $\hat b, u_y \in \mathfrak{g}$,
\begin{subequations}
\begin{align}
\dot E &=-T_{\hat X} R_{X^{-1}} \alpha_{\mathring y}(\hat X, y,  \hat b,u_y,t)-T_I R_{E} \text{Ad}_{\hat X} \tilde b=F_{\mathring y}(E,\tilde b)\\
\dot{\tilde b} &=\beta_{\mathring y}(\hat X, y,\hat b,u_y,t)=H_{\mathring y}(E,\tilde b).
\end{align}
\end{subequations}
It immediately follows that $\alpha_{\mathring y}$ and $\beta_{\mathring y}$ are independent of $u_y$ and $t$. Moreover, since the error $E=\hat X X^{-1}$ is invariant with respect to the transformation $(\hat X,X) \mapsto (\hat X Z, X Z)$ for all $Z\in G$ and the error $\tilde b=\hat b-b$ is invariant with respect to the transformation $(\hat b, b) \mapsto (\hat b+d, b+d)$ for all $d\in \mathfrak{g}$, we have
\begin{subequations}
\begin{align}
\label{error_group_temp1} &-T_{\hat X} R_{X^{-1}} \alpha_{\mathring y}(\hat X, y, \hat b)-T_I R_{E} \text{Ad}_{\hat X} \tilde b=F_{\mathring y}(E,\tilde b)\\
\nonumber &~~~~~~~ =-T_{\hat X Z} R_{(X Z)^{-1}}  \alpha_{\mathring y}(\hat X Z, h(Z,y), \hat b+d)-T_I R_{E} \text{Ad}_{\hat X Z} \tilde b,\\
\label{error_bias_temp1} & \beta_{\mathring y}(\hat X, y,\hat b)=H_{\mathring y}(E,\tilde b)=\beta_{\mathring y}(\hat X Z, h(Z,y),\hat b+d).
\end{align}
\end{subequations}
From (\ref{error_bias_temp1}) it follows that $\beta_{\mathring y}$ is independent of $\hat b$ since the right hand side of (\ref{error_bias_temp1}) depends on $d$ while the left hand side is independent of this variable. This establishes condition \ref{con_autonomy_indep_param} for $\beta_{\mathring y}$. It also follows that $\beta_{\mathring y}$ satisfies the invariance condition $\beta_{\mathring y}(\hat X, y)=\beta_{\mathring y}(\hat X Z, h(Z,y))$ (condition \ref{con_autonomy_beta_invar} in the Theorem). We can rearrange (\ref{error_group_temp1}) to obtain
\begin{align} \label{proof_temp2}
-T_{\hat X} R_{X^{-1}} \alpha_{\mathring y}(\hat X, y, \hat b)&-T_I R_{E} \text{Ad}_{\hat X} \tilde b+T_I R_{E} \text{Ad}_{\hat X Z} \tilde b \\
\nonumber & =-T_{\hat X Z} R_{(X Z)^{-1}}  \alpha_{\mathring y}(\hat X Z, h(Z,y), \hat b+d).
\end{align}
The right hand side of (\ref{proof_temp2}) is a function of $d$ while the left hand side is not. This implies that $\alpha_{\mathring y}$ is independent of $\hat b$ (establishing condition \ref{con_autonomy_indep_param} for $\alpha_{\mathring y}$). We can then rearrange (\ref{proof_temp2}) again to obtain
\begin{align}
\label{proof_temp3} -T_{\hat X} R_{X^{-1}} \alpha_{\mathring y}(\hat X, y)&+T_{\hat X Z} R_{(X Z)^{-1}}  \alpha_{\mathring y}(\hat X Z, h(Z,y))\\
\nonumber &=T_I R_{E} \text{Ad}_{\hat X} \tilde b-T_I R_{E} \text{Ad}_{\hat X Z} \tilde b.
\end{align}
The right hand side of (\ref{proof_temp3}) is a linear function acting on $\tilde b\in \mathfrak{g}$ while the left hand side is completely independent of the variable $\tilde b$. Since $\tilde b$ is arbitrary, this implies that both sides of (\ref{proof_temp3}) are zero. In particular, $T_{\hat X} R_{X^{-1}} \alpha_{\mathring y}(\hat X, y)=T_{\hat X Z} R_{(X Z)^{-1}} \alpha_{\mathring y}(\hat X Z, h(Z,y))$ and $T_I R_{E} \text{Ad}_{\hat X Z} \tilde b=T_I R_{E} \text{Ad}_{\hat X} \tilde b$ for all $\tilde b \in \mathfrak{g}$ and all $E,\hat X,Z \in G$. These equations imply $T_{\hat X} R_Z \alpha_{\mathring y}(\hat X,y)=\alpha_{\mathring y}(\hat X Z,h(Z,y))$ and $\text{Ad}_{Z} \tilde b= \tilde b$ to obtain conditions \ref{con_autonomy_alpha_invar} and \ref{con_autonomy_adj_trivial} imposed in the theorem, respectively. This completes the proof.\carre

\begin{remark} \label{rem_abelian}
If $G$ is a real, finite-dimensional, connected Lie group then condition \ref{con_autonomy_adj_trivial} of Theorem \ref{theo_autonomy} implies that $G$ is Abelian \cite[Proposition 1.91]{knapp2002lie}. By the structure theorem for connected Abelian Lie groups \cite[Proposition III.6.4.11]{bourbaki1989lie}, this means that $G$ is isomorphic to a product $\mathds{R}^n \times (\textsf{S}^1)^m$ where $\mathds{R}^n$ is additive and $(\textsf{S}^1)^m$ denotes the $m$-dimensional torus. This class of Lie groups is far more specific than the Lie groups that are encountered in many practical applications. For robotics applications, the Lie groups typically considered are $\textsf{SO}(3)$ and $\textsf{SE}(3)$ both of which are non-Abelian. Theorem \ref{theo_autonomy} in particular implies that all implementable geometric bias estimators on $\textsf{SO}(3)$ and $\textsf{SE}(3)$ proposed in the literature produce non-autonomous standard error dynamics (see \cite{Mahony08} and \cite{vasconcelos2010nonlinear}).
\carrew
\end{remark}

\section{Observer Design and Analysis} \label{sec_observer_design}
We propose the following implementable group estimator,
\begin{align} \label{observer_left_dyn}
\dot {\hat X}=\hat X [u_y-\hat b]- K_{\mathring y}(\hat X, y,\hat b,u_y,t) [\text{D}_1 \phi_{\mathring y}(\hat X,y)],
\end{align}
with $\hat X(t_0)=\hat X_0$ where $\phi_{\mathring y}\colon G\times M \to \mathds{R}^+$ is a $C^2$ cost function, $\text{D}_1 \phi_{\mathring y} (\hat X,y) \in T_{\hat X}^* G$ denotes the differential of $\phi_{\mathring y_i}$ with respect to its first argument evaluated at the point $(\hat X,y)$ and $K_{\mathring y}(\hat X, y,\hat b,u_y,t)$ is a linear \emph{gain mapping} from $T^*_{\hat X} G$ to $ T_{\hat X} G$. Note that $\mathring y$ is considered to be a parameter for $K_{\mathring y}$ and $\phi_{\mathring y}$. The above group estimator matches the structure of (\ref{group_estimator}) where the innovation $\alpha_{\mathring y}$ is generated by applying the gain mapping $K_{\mathring y}$ to the differential $\text{D}_1 \phi_{\mathring y}$. By Theorem \ref{theo_autonomy}, we already know that the above estimator cannot produce autonomous error dynamics for a general Lie group. Hence, there is no reason to omit the arguments $\hat b,u_y$ and $t$ of the gain mapping. If the gain mapping $K_{\mathring y}$ is symmetric positive definite and independent of $\hat b,u_y$ and $t$, the above group estimator simplifies to the gradient-like observers proposed in \cite{lageman2010gradient} for the bias free case, and in \cite{khosravian2013bias} for the case including bias.

We consider the following bias estimator,
\begin{align} \label{b_hat_dy}
\dot{\hat b}= \Gamma \circ T_I L_{\hat X}^* [\text{D}_1 \phi_{\mathring y} (\hat X, y)],~~~~~\hat b(t_0)=\hat b_0,
\end{align}
where $T_I L_{\hat X}^*\colon T^*_{\hat X} G \to \mathfrak{g}^*$ is the dual of the map $T_I L_{\hat X}$ (see Section \ref{preliminaries}) and $\Gamma\colon \mathfrak{g}^* \to \mathfrak{g}$ is a constant gain mapping.

We will require the following assumptions for statement of results.
\begin{itemize}
\labitem{(A1)}{assumption_matrix}{Lie group $G$ has a faithful representation as a finite-dimensional matrix Lie group. That is, there exist a positive integer $m$ and an injective Lie group homomorphism $\Phi: G \to \textsf{GL}(m)$ into the group $\textsf{GL}(m)$ of invertible $m \times m$ matrices. Note that $\Phi(G)$ is a matrix subgroup of $\textsf{GL}(m)$.}
\labitem{(A2)}{assumption_bounded}{[\textit{boundedness conditions}]~~$\Phi(X)$, $\Phi(X)^{-1}$, $u$ and $K(\hat X,y,\hat b,u_y,t)$ are bounded along the system trajectories.}
\labitem{(A3)}{assumption_differentiable}{[\textit{differentiability conditions}]~~$\dot u(t)$, the first differentials of $h_i(X,\mathring y_i)$ and  $K(\hat X,y,\hat b,u_y,t)$, as well as the first and the second differential\footnote{Second differential of $\phi$ is either in the sense of embedding the Lie group into $\mathds{R}^{m \times m}$ or in the sense of employing a Riemannian metric.} of $\phi(\hat X,y)$ with respect to all of their arguments are bounded along the system trajectories.}
\end{itemize}

\begin{theorem} \label{theo_observer_stability}
Consider the observer (\ref{observer_left_dyn})-(\ref{b_hat_dy}) for the system (\ref{group_left_dyn})-(\ref{u_y}). Suppose that assumptions \ref{assumption_matrix}, \ref{assumption_bounded} and \ref{assumption_differentiable} hold. Assume moreover that the gain mappings $K$ and $\Gamma$, and the cost function $\phi$ satisfy the following properties;
\begin{enumerate}
\labitem{(a)}{con_obsstability_K}{The gain mapping $K_{\mathring y}\!\colon\! T^*_{\hat X} G \to T_{\hat X} G$ is uniformly positive definite (not necessarily symmetric). That is, there exist positive constants $\underline{k}$ and $\overline{k}$ and a continues vector norm $|.|$ on $T^*_{\hat X} G$ such that for all $v^*\in T^*_{\hat X} G$ we have $\underline{k} |v^*|^2 \le v^*[K_{\mathring y}(\hat X,y,\hat b,u_y,t)[v^*]] \!\le\! \overline{k} |v^*|^2$.}
\labitem{(b)}{con_obsstability_Gamma}{The gain mapping $\Gamma\!\colon\! \mathfrak{g}^* \!\!\to\!\! \mathfrak{g}$ is symmetric positive definite.}
\labitem{(c)}{con_obsstability_phi_invar}{The cost $\phi_{\mathring y}$ is right invariant, that is $\phi_{\mathring y} (\hat X Z, h(Z,y)) = \phi(\hat X, y)$ for all $\hat X,Z \in G$ and all $y \in M$.}
\labitem{(d)}{con_obsstability_phi_PD}{The cost $\phi_{\mathring y}(.,\mathring y)\colon G\to \mathds{R}^+,~E \mapsto \phi_{\mathring y}(E,\mathring y)$ is locally positive definite around $E=I$ and it has an isolated critical point at $E=I$.}
\end{enumerate}
Then the error dynamics $(\dot E,\dot{\tilde b})$ is uniformly locally asymptotically stable at $(I,0)$. \carrew
\end{theorem}

\textit{Proof: }
The following result is used in the development later in this proof.
\begin{lemma} \label{lemma_Dphi_simplify}
Let $\phi_{\mathring y}\colon G \times M \to \mathds{R}^+$ be a right-invariant cost function in the sense defined in part \ref{con_obsstability_phi_invar} of Theorem \ref{theo_observer_stability}. Then we have
\begin{align}
\label{D_phi_E2X} \text{D}_1 \phi_{\mathring y} (\hat X, y) &=T_{\hat X} R_{X^{-1}}^* [\text{D}_1 \phi_{\mathring y} (E,\mathring y)]\\
\label{D_phi_X2E} \text{D}_1 \phi_{\mathring y} (E,\mathring y) &=\text{D}_1 \phi_{\mathring y} (\hat X, y) \circ T_{E} R_X
\end{align} %\carrew
%=\text{D}_1 \phi_{\mathring y} (E,\mathring y) \circ T_{\hat X} R_{X^{-1}}
\end{lemma}

Proof of Lemma \ref{lemma_Dphi_simplify} is given in appendix. For simplicity, we denote $K_{\mathring y}(\hat X,y,\hat b,u_y,t)$ by $K_{\mathring y}(.)$. Considering (\ref{group_left_dyn}), (\ref{u_y}), and (\ref{observer_left_dyn}), the group error dynamics are given by
\begin{align}
\nonumber & \dot E=\dot{\hat X} X^{-1}+\hat X \dot{X^{-1}}=\!T_{\hat X} R_{X^{-1}} \!\circ\! T_I L_{\hat X} [u] \!-\! T_{\hat X} R_{X^{-1}} \!\circ\! T_I L_{\hat X} [\tilde b]\\
\nonumber &~~~~~~~~~~~~~~~~~~~~~~~-\!  T_{\hat X} R_{X^{-1}} \!\circ\! K_{\mathring y}(.)[\text{D}_1 \phi_{\mathring y}(\hat X,y)] \!-\!T_{\hat X} R_{X^{-1}} \!\circ\!  T_I L_{\hat X} [u]\\
\label{group_error_dyn} &=\!\!-T_{\!\hat X} R_{\! X^{-1}} \!\!\circ\! T_{\!I} L_{\!\hat X} [\tilde b]\!-\!  T_{\!\hat X} R_{X^{-1}} \!\circ\! K_{\mathring y}(.) \!\circ\! T_{\!\hat X} R_{\!X^{-1}}^* [\text{D}_1 \phi_{\mathring y} (E,\mathring y)],
\end{align}
where $E$ is as in (\ref{right_error}) and (\ref{D_phi_E2X}) is used in the last line of (\ref{group_error_dyn}). Now, consider the candidate Lyapunov function,
\begin{align}
\mathcal{L}(E,\tilde b)=\phi_{\mathring y}(E,\mathring y)+\frac{1}{2}\Gamma^{-1}[\tilde b][\tilde b].
\end{align}
The Lyapunov candidate is at least locally positive definite due to conditions \ref{con_obsstability_Gamma} and \ref{con_obsstability_phi_PD}. The time derivative of $\mathcal{L}$ is given by
\begin{align} \label{dot_L_tilde_b}
\dot{\mathcal{L}}(E,\tilde b)=\text{D}_1\phi_{\mathring y}(E,\mathring y) [\dot E] +\Gamma^{-1}[\dot{\tilde b}][\tilde b] .
\end{align}
Recalling that $\dot{\tilde b}=\dot{\hat b}$ and substituting $\dot E$ form (\ref{group_error_dyn}) in (\ref{dot_L_tilde_b}), we obtain
\begin{align}
\nonumber \dot{\mathcal{L}}(E,\tilde b)\!=\!&-\!\text{D}_1\phi_{\mathring y}(E,\mathring y) [T_{\!\hat X} R_{\!X^{\!-1}} \!\circ\! K_{\mathring y}(.) \!\circ\! T_{\hat X} R_{X^{-1}}^* [\text{D}_1 \phi_{\mathring y} (E,\mathring y)]] \\
\label{dot_L_temp1} &-\text{D}_1\phi_{\mathring y}(E,\mathring y)[T_{\hat X} R_{X^{-1}} \!\circ\! T_I L_{\hat X} [\tilde b]]+\Gamma^{-1}[\dot{\hat b}][\tilde b].
\end{align}
Using (\ref{D_phi_X2E}), we conclude
\begin{align}
\nonumber \dot{\mathcal{L}}(E,\tilde b)\!=\!&-\!\text{D}_1\phi_{\mathring y}(E,\mathring y)[T_{\!\hat X} R_{\!X^{\!-1}} \!\circ\! K_{\mathring y}(.) \!\circ\! T_{\hat X} R_{X^{-1}}^* [\text{D}_1 \phi_{\mathring y} (E,\mathring y)]]  \\
\nonumber &-\!\text{D}_1 \phi_{\mathring y} (\hat X, y) \!\circ\! T_{\!E} R_X \!\circ\! T_{\hat X} R_{X^{-1}} \circ T_I L_{\hat X} [\tilde b]+\Gamma^{-1}[\dot{\hat b}][\tilde b]\\
\nonumber =&-\!\text{D}_1\phi_{\mathring y}(E,\mathring y)[T_{\!\hat X} R_{\!X^{\!-1}} \!\circ\! K_{\mathring y}(.) \!\circ\! T_{\hat X} R_{X^{-1}}^* [\text{D}_1 \phi_{\mathring y} (E,\mathring y)]]\\
\label{dot_L_temp1b} &-\text{D}_1 \phi_{\mathring y} (\hat X, y) \circ T_I L_{\hat X} [\tilde b]+\Gamma^{-1}[\dot{\hat b}][\tilde b].
\end{align}
Now, replacing $\dot{\hat b}$ with (\ref{b_hat_dy}) we obtain
\begin{align}
\nonumber \dot{\mathcal{L}}&(E,\tilde b) \!=\!-\text{D}_1\phi_{\mathring y}(E,\mathring y)[T_{\!\hat X} R_{\!X^{\!-1}} \!\circ\! K_{\mathring y}(.) \circ T_{\hat X} R_{X^{-1}}^* [\text{D}_1 \phi_{\mathring y} (E,\mathring y)]]\\
\label{dot_L_temp2} &-\!\text{D}_1 \phi_{\mathring y} (\hat X,\! y) \!\circ\! T_{\!I} L_{\hat X} [\tilde b]\!+\!\Gamma^{-1} \!\!\circ\! \Gamma \!\circ\! T_I L_{\hat X}^* \!\circ\! \text{D}_1 \phi_{\mathring y} (\hat X, y)[\tilde b].
\end{align}
Duality implies $\text{D}_1 \phi_{\mathring y} (\hat X, y) \circ T_I L_{\hat X}= T_I L_{\hat X}^* \circ \text{D}_1 \phi_{\mathring y} (\hat X, y)$ and (\ref{dot_L_temp2}) simplifies to
\begin{align} \label{dot_L_temp3}
\dot{\mathcal{L}}(E,\tilde b)&\!=\!-\text{D}_1\phi_{\mathring y}(E,\!\mathring y)[T_{\!\hat X} R_{\!X^{\!-1}} \!\!\circ\! K_{\mathring y}(.) \!\circ\! T_{\!\hat X} R_{\!X^{-1}}^* [\text{D}_1 \phi_{\mathring y} (E,\!\mathring y)]].
\end{align}
Since $K_{\mathring y}(.)$ is assumed to be positive definite and the map $T_{\hat X} R_{X^{-1}}$ is full rank, the map $T_{\hat X} R_{X^{-1}} \circ K_{\mathring y}(.) \circ T_{\hat X} R_{X^{-1}}^* $ is positive definite. This implies that $\dot{\mathcal{L}}(E,\tilde b) \le 0$ and hence the Lyapunov function is non-increasing along the system trajectories. We adopt the proof of \cite[Theorem 4.8]{Khalil} to prove uniformly local stability of error dynamics. Recalling assumption \ref{assumption_matrix}, distance to the identity element of $G$ is denoted by $d(.)$ and is induced by Frobenius norm on $\Phi(G) \subset \mathds{R}^{m\times m}$ via $d(E):=\|\text{Id}-\Phi(E)\|_F$ where $\text{Id}$ is the identity matrix. Define the compound error $\tilde x:=(E,\tilde b)\in G \times \mathfrak{g}$ and obtain the distance of $\tilde x$ to $(I,0)$ by $l(\tilde x)^2:=d(E)^2+\|\tilde b\|_{\mathfrak{g}}^2$ where $\|.\|_{\mathfrak{g}}$ denotes a norm on $\mathfrak{g}$. Using assumption \ref{con_obsstability_phi_PD}, there exist a ball $B_r:=\{E\in G:~d(E) \le r \}$ such that $\phi_{\mathring y}(.,\mathring y)$ is positive definite on $B_r$. Consequently $\mathcal{L}(\tilde x)$ is positive definite on $\bar B_r:=\{\tilde x\in G\times \mathfrak{g}:~ l(\tilde x) \le r \}$. Choose $c < \text{min}_{l(\tilde x)=r}\mathcal{L}(\tilde x)$ and define $\Omega_c:=\{\tilde x \in \bar B_r|~ \mathcal{L}(\tilde x) \le c \}$. Since $\dot{\mathcal{L}}(t) \le 0$, any solution starting in $\Omega_c$ at $t_0$ remains in $\Omega_c$ for all $t \ge t_0$. On the other hand, since $\mathcal{L}(\tilde x)$ is positive definite on $\Omega_c \subset \bar B_r$, there exist class $\mathcal{K}$ functions $\eta_1$ and $\eta_2$ such that $\eta_1(l(\tilde x)) \le \mathcal{L}(\tilde x) \le \eta_2(l(\tilde x))$ for all $\tilde x \in \Omega_c$ \cite[Lemma 4.3]{Khalil}. Consequently, we have $l(\tilde x(t)) \le \eta_1^{-1}(\mathcal{L}(\tilde x(t))) \le \eta_1^{-1}(\mathcal{L}(\tilde x(t_0))) \le \eta_1^{-1}(\eta_2(l(\tilde x(t_0))))$ which implies $l(\tilde x(t)) \le \eta_1^{-1} \circ \eta_2 (l(\tilde x(t_0))) $. Since $\eta_1^{-1} \circ \eta_2$ is a class $\mathcal{K}$ function (by \cite[Lemma 4.2]{Khalil}), the equilibrium point $\tilde x=(I,0)$ is uniformly stable for all initial conditions starting in $\Omega_c$ \cite[Lemma 4.5]{Khalil}. Moreover, the error $E$ is bounded by $d(E(t)) \le l(\tilde x(t)) \le \eta_1^{-1}(\mathcal{L}(\tilde x(t_0))) \le \eta_1^{-1}(c)$ for such initial conditions.

Boundedness of $\tilde x(t)$ implies that $E(t)$ and $\tilde b(t)$ are bounded with respect to $d(.)$ and $\|.\|_\mathfrak{g}$, respectively. Differentiating (\ref{dot_L_temp3}) with respect to time and considering the boundedness of $(E(t),\tilde b(t))$ together with assumptions \ref{assumption_bounded} and \ref{assumption_differentiable}, one can conclude that $\ddot{\mathcal{L}}(t)$ is bounded and hence $\dot{\mathcal{L}}(t)$ is uniformly continuous. By invoking Barbalat's lemma we conclude that $\dot{\mathcal{L}}(t) \to 0$. This together with condition \ref{con_obsstability_K} implies that $\text{D}_1 \phi_{\mathring y} (E(t),\mathring y) \to 0$. Since $\phi_{\mathring y}(E,I)$ has an isolated critical point at $E=I$, there exist a ball $B_{\bar c} \subset G$ such that $E=I$ is the only point in $B_{\bar c}$ where $\text{D}_1 \phi_{\mathring y}(.,\mathring y)$ is zero. We proved before that $E(t) \in B_{\eta_1^{-1}(c)}$ for all initial conditions starting in $\Omega_c$. Choosing $c < \text{min}(\eta_1 (\bar c), \text{min}_{l(\tilde x)=r}\mathcal{L}(\tilde x) )$ ensures that $E=I$ is the only critical point in $B_{\eta_1^{-1}(c)}$. This implies that $E(t) \to I$ for all initial conditions starting in $\Omega_c$. Using (\ref{group_left_dyn}), (\ref{group_estimator}), and (\ref{group_error_dyn}), recalling assumptions \ref{assumption_bounded} and \ref{assumption_differentiable}, and using a local coordinate representation, one can verify that $\ddot E(t)$ is bounded and hence $\dot E(t)$ is uniformly continuous. Thus, by invoking Barbalat's lemma we have $\dot E(t) \to 0$. Considering $E(t),\dot E(t) \to 0$ together with error dynamics (\ref{group_error_dyn}) implies that $\tilde b(t) \to 0$ for all initial errors starting in $\Omega_c$. This completes the proof of uniformly local asymptotic stability of the error dynamics.\carre

The following theorem proposes additional conditions to guarantee local \emph{exponential} stability of the error dynamics.

\begin{theorem} \label{theo_observer_exp_stability}
Consider the observer (\ref{observer_left_dyn})-(\ref{b_hat_dy}) for the system (\ref{group_left_dyn})-(\ref{u_y}). Suppose that assumptions \ref{assumption_matrix} and \ref{assumption_bounded} and conditions \ref{con_obsstability_K}, \ref{con_obsstability_Gamma} and \ref{con_obsstability_phi_invar} of Theorem  \ref{theo_observer_stability} hold. Assume moreover that;
\begin{enumerate}
\labitem{(d)}{con_observer_exp_phi}{$\text{D}_1 \phi_{\mathring y}(I,\mathring y)=0$ and $\text{Hess}_1 \phi_{\mathring y}(I,\mathring y): \mathfrak{g} \to \mathfrak{g}^*$ is symmetric positive definite.}
\labitem{(e)}{con_observer_exp_X_cond}{The condition number of $\Phi(X(t))$ is  bounded for all $t \ge t_0$ (uniformly in $t_0$).}
\end{enumerate}
Then, the error dynamics $(E(t),\tilde b(t))$ is uniformly locally exponentially stable at $(I,0)$.
\carrew
\end{theorem}

\textit{Proof: }
The group error dynamics (\ref{group_error_dyn}) can be rewritten as
\begin{align}
 \label{closed_loop_error_E_b} \dot E \!=\!-\!  T_{\!\hat X} R_{\!X^{\!-1}} \!\!\circ\! K_{\mathring y}(\hat X,\!y,\!\hat b,\!u_y,\!t) \!\circ\!\! T_{\!\hat X} R_{\! X^{\!-1}}^* [\text{D}_1 \phi_{\mathring y} (E,\mathring y)]\! -\!T_{\!I} L_{E} \text{Ad}_{\!X} \![\tilde b].
\end{align}
Using (\ref{D_phi_E2X}) and (\ref{b_hat_dy}), the bias error dynamics is obtain as
\begin{align}
\nonumber \dot{\tilde b}&= \Gamma \circ T_I L_{\hat X}^* \circ T_{\hat X} R_{X^{-1}}^* [\text{D}_1 \phi_{\mathring y} (E,\mathring y)]\\
 \label{closed_loop_error_b_b} &=\Gamma \circ \text{Ad}_{X}^* \circ T_I L_{E}^* [\text{D}_1 \phi_{\mathring y}(E,\mathring y)].
\end{align}
%\begin{align}
%\label{closed_loop_error_b_b} \!\!\!\!\!\!\!\dot{\tilde b}\!=\! \Gamma \!\circ T_{\!I} L_{\hat X}^* \!\circ\! T_{\!\hat X} R_{\!X^{-1}}^* [\text{D}_1 \phi_{\mathring y} (E,\!\mathring y)]\!=\!\Gamma \!\circ\! \text{Ad}_{X}^* \!\circ\! T_{\!I} L_{E}^* [\text{D}_1 \phi_{\mathring y}(E,\!\mathring y)].
%\end{align}
Defining $\epsilon,\delta\in\mathfrak{g}$ as the first order approximation of $E$ and $\tilde b$ respectively, linearizing the error dynamics (\ref{closed_loop_error_E_b})-(\ref{closed_loop_error_b_b}) around $(I,0)$ and neglecting all terms of quadratic or higher order in $(\epsilon,\delta)$ yields
\begin{align}
\nonumber \dot \epsilon\!&=\! -  T_{\!X} R_{\!X^{\!-1}} \!\circ\! K_{\mathring y}(X\!,\!y,\!b,\!u_y,\!t) \!\circ\! T_{\!X} R_{\!X^{\!-1}}^* \!\circ\! \text{Hess}_1\phi_{\mathring y}(I,\!\mathring y) [\epsilon]\\
\label{linear_error_E_b} &~~~-\!\text{Ad}_{X} \![\delta],\\
\label{linear_error_b_b} \dot\delta&=\Gamma \circ \text{Ad}_{X}^* \circ \text{Hess}_1\phi_{\mathring y}(I,\mathring y) [\epsilon],
\end{align}
where $\text{Hess}_1\phi_{\mathring y} (I,\mathring y)\colon \mathfrak{g} \to \mathfrak{g}^*$ denotes the Hessian operator which is intrinsically defined at the critical point of the cost \cite{absil2009optimization}. In order to investigate the stability of the linearized error dynamics, we consider a basis for the involved tangent spaces and rewrite (\ref{linear_error_E_b})-(\ref{linear_error_b_b}) in matrix format. To this end, consider a basis $\{\text{e}_j \}$ for $\mathfrak{g}$ and its corresponding dual basis for $\mathfrak{g}^*$. Obtain the basis $\{ \text{e}_j X \}$ for the vector space $T_{X} G$ by right translating $\{\text{e}_j\}$ and consider its corresponding dual basis $\{(\text{e}_j X)^* \}$ for $T_{X}^* G$. Denote by $[\![\epsilon ]\!],[\![\delta ]\!]$ the representation of the vectors $\epsilon,\delta$ with respect to the basis $\{ \text{e}_j \}$. Denote the matrix representation of the maps $K_{\mathring y}(X,y,b,u_y,t) \colon T^*_{X} G \to T_{X} G$, $\Gamma\colon \mathfrak{g}^* \to \mathfrak{g}$, $\text{Hess}_1\phi_{\mathring y}(I,\mathring y)\colon \mathfrak{g} \to \mathfrak{g}^*$ and $\text{Ad}_{X}\colon \mathfrak{g} \to \mathfrak{g}$ with respect to the above bases for their corresponding domain and co-domain by $[\![ K ]\!]$, $[\![ \Gamma ]\!]$, $[\![ H ]\!]$ and $[\![ \text{Ad}_{X} ]\!]$ respectively. Note that the matrix representation of $T_{\hat X} R_{X^{-1}}\colon  T_{X} G \to \mathfrak{g}$ with respect to the corresponding basis for its domain and co-domain is the identity matrix. Hence, the matrix representation of the error dynamics (\ref{linear_error_E_b})-(\ref{linear_error_b_b}) is obtained as
\begin{align} \label{linear_error_dyn_matrix}
\left[ {\begin{array}{*{20}{c}}
{\dot{[\![ \epsilon ]\!]}}\\
{\dot{[\![ \delta ]\!]}}
\end{array}} \right]
= \left[ {\begin{array}{*{20}{c}}
{- [\![ K ]\!] [\![ H ]\!] }&{-[\![\text{Ad}_{X}]\!]}\\
{[\![ \Gamma ]\!] [\![ \text{Ad}_{ X}]\!]^\top [\![ H ]\!]}&{0}
\end{array}} \right]
\left[ {\begin{array}{*{20}{c}}
{[\![ \epsilon ]\!]}\\
{[\![ \delta ]\!] }
\end{array}} \right].
\end{align}

Since $\Gamma$ is symmetric positive definite, there exists a full rank square matrix $L$ such that $[\![\Gamma ]\!]=L^\top L$. Consider the change of coordinates $\bar \epsilon:= L [\![ \epsilon ]\!]$ and $\bar \delta:= L^{-\top} [\![ \delta ]\!]$. Using (\ref{linear_error_dyn_matrix}), the dynamics of the new error coordinates are obtained as
\begin{align} \label{linear_error_dyn_matrix_new}
\left[ {\begin{array}{*{20}{c}}
{ \dot{\bar{\epsilon}}}\\
{\dot{\bar{\delta}}}
\end{array}} \right]
= \left[ {\begin{array}{*{20}{c}}
{- L [\![ K ]\!] [\![ H ]\!] L^{-1} }&{-L [\![\text{Ad}_{X}]\!] L^\top}\\
{ L [\![ \text{Ad}_{X}]\!]^\top
[\![ H ]\!] L^{-1}}&{0}
\end{array}} \right]
\left[ {\begin{array}{*{20}{c}}
{\bar{\epsilon}}\\
{\bar{\delta}}
\end{array}} \right].
\end{align}
Consider initial conditions $X(t_0)$ for system (\ref{group_left_dyn}) and $(\hat X(t_0),\hat b(t_0))$ for the estimator (\ref{observer_left_dyn})-(\ref{b_hat_dy}), respectively. Introducing the parameter $\lambda=(t_0,X(t_0),\hat X(t_0),\hat b(t_0))\in \mathcal{D}$ where
$\mathcal{D}:=\mathds{R}\times G \times G \times \mathfrak{g}$, the trajectories of $X, \hat X, \hat b$ and $y$ can be viewed as functions of $t$ and $\lambda$. Define $A(t,\lambda):=-  L [\![ K ]\!] [\![ H ]\!] L^{-1}$, $B(t,\lambda):=-L [\![\text{Ad}_{X}]\!]^\top L^\top$, and $P:=L^{-\top}[\![ H ]\!] L^{-1}$. The system (\ref{linear_error_dyn_matrix_new}) belongs to the following standard class of parameterized linear time-varying systems discussed extensively in the literature \cite{morgan1977uniform,morgan1977stability,loria2002uniform}.
\begin{align} \label{linearzed_dy_structure}
\left[ {\begin{array}{*{20}{c}}
{\dot{\bar \epsilon}}\\
{\dot{\bar \delta}}
\end{array}} \right]
= \left[ {\begin{array}{*{20}{c}}
{A(t,\lambda) }&{B(t,\lambda)^\top}\\
{-B(t,\lambda) P}&{0}
\end{array}} \right]
\left[ {\begin{array}{*{20}{c}}
{\bar \epsilon}\\
{\bar \delta}
\end{array}} \right]
\end{align}
We can now verify the conditions of \cite[Theorem 1]{loria2002uniform} to prove the stability of system (\ref{linear_error_dyn_matrix_new}). Both $B(t,\lambda)$ and its time derivative are bounded due to Assumption \ref{assumption_bounded}. Since $\text{Hess}_1 \phi_{\mathring y} (I,\mathring y)$ is symmetric positive definite and $L$ has full rank, the matrix $P$ is symmetric positive definite and it is bounded by $\underline{\sigma}(H) \underline{\sigma}(L)^{-2} I \le P \le \bar{\sigma}(H) \bar{\sigma}(L)^{-2} I$ where $\underline{\sigma}(.)$ and $\bar{\sigma}(.)$ denote the smallest and largest singular value of a matrix respectively. Define $-Q:=\dot P+A(t,\lambda)^\top P+P A(t,\lambda)=-([\![ H ]\!] L^{-1})^\top ([\![K]\!]^\top + [\![K]\!])([\![ H ]\!] L^{-1})$. Using condition \ref{con_obsstability_K} of Theorem \ref{theo_observer_stability} and recalling assumption \ref{assumption_bounded}, there exist positive constants $k_1$ and $k_2$ such that $2 k_1 \text{Id} \le  [\![K]\!]^\top \!\!+ [\![K]\!] \le 2 k_2 \text{Id}$ where $\text{Id}$ is the identity matrix. This ensures that $Q$ is uniformly symmetric positive definite and we have $2 k_1 \underline{\sigma}(H)^2 \underline{\sigma}(L)^{-2} \text{Id} \le Q \le 2\bar{\sigma}(H)^2 k_2 \bar{\sigma}(L)^{-2} \text{Id}$. It only remains to investigate whether $B(t,\lambda)$ is $\lambda$-uniformly persistently exciting \cite[equation (10)]{loria2002uniform}. Embed the Lie algebra $\mathfrak{g}$ into $\mathds{R}^{m \times m}$. Invoking the property $\text{vec}(\Phi(X) w \Phi(X)^{-1})=\Phi(X)^{-\top} \otimes \Phi(X) \text{vec}(w)$ where $\text{vec}(w) \in \mathds{R}^{m^2}$ is the vectorization of the matrix $w\in \mathfrak{g}$ and $\otimes$ denotes the Kronecker product, one can conclude that the matrix representation of $\text{Ad}_{X}:\mathds{R}^{m \times m} \to \mathds{R}^{m \times m}$ with respect to the standard basis for its domain and co-domain is given by $[\![ \text{Ad}_{X}]\!]=\Phi(X)^{-\top} \otimes \Phi(X)$. Thus $\underline{\sigma}([\![ \text{Ad}_{X}]\!])=\underline{\sigma}(\Phi(X)^{-\top})\underline{\sigma}(\Phi(X))=\text{cond}(\Phi(X))^{-1}$ where $\text{cond}(\Phi(X))$ denotes the condition number of $\Phi(X) \in \textsf{GL}(m)$. Since $\mathfrak{g} \subset \mathds{R}^{m \times m}$, the minimum singular value of $\text{Ad}_X:\mathfrak{g} \to \mathfrak{g}$ is larger than or equal to the minimum singular value of $\text{Ad}_X:\mathds{R}^{m \times m} \to \mathds{R}^{m \times m}$. Using condition \ref{con_observer_exp_X_cond}, there exists a positive constant $c_0$ such that $\text{cond}(\Phi(X)(t)) \le c_0$. Hence, $\underline{\sigma}(B(t,\lambda)B(t,\lambda)^\top)= \underline{\sigma}(L [\![\text{Ad}_{X}]\!]^\top [\![\Gamma]\!] [\![\text{Ad}_{X}]\!] L^\top) \ge \underline{\sigma}(L)^2 \underline{\sigma}([\![\Gamma]\!]) c_0^{-2}:=\bar{c}_0$.
 Integrating both sides yields $\int_{\tau}^{t+T} {B(\tau,\lambda)B(\tau,\lambda)^\top}d\tau \ge \bar{c}_0T \text{Id}$ which completes the requirements of \cite[Theorem 1]{loria2002uniform}. Hence, the equilibrium $(0,0)$ of the  (\ref{linear_error_dyn_matrix_new}) is uniformly exponentially stable. This implies that the equilibrium $(0,0)$ of the linearized system (\ref{linear_error_dyn_matrix_new}) is uniformly exponentially stable and consequently the equilibrium $(I,0)$ of the nonlinear error dynamics (\ref{closed_loop_error_E_b})-(\ref{closed_loop_error_b_b}) is uniformly locally exponentially stable \cite[Theorem 4.15]{Khalil} (note that what is referred to as uniform exponential stability here is the same as exponential stability in the sense of \cite{Khalil}).

Owing to the parameter-dependent analysis, the obtained exponential stability is
uniform with respect to the choice of all initial conditions in $\lambda$
and not only with respect to the choice of $E(t_0)$ and $\tilde{b}(t_0)$ for a given $\hat{X}$.\carre

\begin{remark} \label{rem_matrix_vs_general}
For the stability analysis, we assume that $G$ allows a matrix Lie group representation (by assumption \ref{assumption_matrix}). Nevertheless, the actual formulas of the proposed observer (\ref{observer_left_dyn})-(\ref{b_hat_dy}) can be computed without requiring any matrix structure for the Lie group, owing to the representation-free formulation of the proposed observer. We only require the matrix Lie group representation of $G$ to interpret the boundedness conditions on $\Phi(X)$, $\Phi(X^{-1})$, and $\text{cond}(\Phi(X))$. We will illustrate this point further with an example in section \ref{sec_SE3_exmple}. Boundedness of $\Phi(X(t))$ and $\Phi(X^{-1}(t))$ are usually mild conditions in practice. Moreover, it is easy to verify that $\text{cond}(X(t))$ is bounded (uniformly in $t_0$) if $\Phi(X(t))$ and $\Phi(X^{-1}(t))$ are bounded (uniformly in $t_0$). For the special case where the considered Lie group is $\textsf{SO}(3)$, all of these boundedness conditions are satisfied automatically since we have $\|\Phi(X(t))\|_F^2=\text{tr}(\Phi(X)^\top \Phi(X))=\text{tr}(I_{3 \times 3})=3$ for all $X \in \textsf{SO}(3)$. In section \ref{sec_SE3_exmple}, we interpret the boundedness requirements for the Lie group $\textsf{SE}(3)$ as well.\carrew
\end{remark}

It is possible to replace the requirement for boundedness of $\Phi(X(t))$, $\Phi(X^{-1}(t))$, and $\text{cond}(\Phi(X(t)))$ respectively with the boundedness of $\Phi(\hat X(t))$, $\Phi(\hat X^{-1}(t))$, and $\text{cond}(\Phi(\hat X(t)))$ in Theorems \ref{theo_observer_stability} and \ref{theo_observer_exp_stability} and still prove the same stability results. Boundedness conditions on $\hat X$ are always verifiable in practice.

Theorem \ref{theo_observer_stability} does not necessarily require a \emph{symmetric} gain mapping $K_{\mathring y}$. Also, we do not impose any invariance condition on this gain mapping. Condition \ref{con_obsstability_K} of Theorem \ref{theo_observer_stability} only requires the symmetric part of $K_{\mathring y}$, denoted by $K^s_{\mathring y}$, to be uniformly positive definite. Considering a basis for $T_{\hat X} G$ and the corresponding dual basis for $T^*_{\hat X} G$, condition \ref{con_obsstability_K} of Theorem \ref{theo_observer_stability} implies that the matrix representation of $K^s_{\mathring y}(.)\colon T^*_{\hat X}G \to T_{\hat X} G$ with respect to these bases is uniformly symmetric positive definite. In practice, we will use this property to design a suitable gain mapping and obtain the innovation term of the observer. We will illustrate this method with an example in Section \ref{sec_SE3_exmple}.

Condition \ref{con_obsstability_phi_PD} of Theorem \ref{theo_observer_stability} is milder than condition \ref{con_observer_exp_phi} of Theorem \ref{theo_observer_exp_stability} or similar conditions imposed in \cite{lageman2010gradient} and \cite{khosravian2013bias}. This allows the choice of much larger class of cost functions to generate innovation terms that guarantee the asymptotic stability of error dynamics.

In the special case where $K_{\mathring y}$ is uniformly symmetric positive definite and is independent of the arguments $\hat b,u_y$ and $t$, the term $K_{\mathring y}(\hat X,y) [\text{D}_1 \phi_{\mathring y}(\hat X,y)]$ simplifies to $\text{grad}_1 \phi_{\mathring y}(\hat X,y)$ where $\text{grad}_1$ denotes the gradient with respect to the Riemannian metric on $G$ induced by the gain mapping. In this case, the observer (\ref{observer_left_dyn})-(\ref{b_hat_dy}) simplifies to the gradient-like observer discussed in \cite[equations (7)-(8)]{khosravian2013bias} where the gain mapping $\Gamma$ is a scalar, or the observer of \cite{lageman2010gradient} for the bias-free case. If in addition we assume that $K_{\mathring y}$ satisfies the invariance condition $T_{\hat X} R_Z \circ K_{\mathring y} (\hat X, y) \circ T_{\hat X} R_Z^*=K_{\mathring y}(\hat X Z,h(Z,y))$, the induced Riemannian metric on $G$ would be right-invariant. In this case, the error dynamics (\ref{closed_loop_error_E_b})-(\ref{closed_loop_error_b_b}) correspond to the perturbed gradient-like error dynamics given by \cite[equations (17)-(18)]{khosravian2013bias}. The larger class of gain mappings together with the larger class of cost functions proposed in this paper ensures that the proposed observer allows a much larger class of observers comparing to the authors' previous work \cite{khosravian2013bias,lageman2008observer,lageman2010gradient}. The discussion presented here shows also that a non-invariant Riemannian metric can be employed for the bias-free case to design the innovation term of the gradient-like observers in \cite{lageman2008observer,lageman2010gradient,khosravian2013bias}. In this case, the resulting error dynamics would be stable as long as the conditions on the cost function are satisfied, but the error dynamics would be non-autonomous. Non-invariant gains also lead to observers that are not symmetry-preserving in the sense of \cite{bonnabel2009non}.

\section{Constructing Invariant Cost Functions on Lie Groups} \label{sec_realizability_grad}
In Section \ref{sec_observer_design}, we propose the observer (\ref{observer_left_dyn})-(\ref{b_hat_dy}) that depend on the differential of the cost function $\phi_{\mathring y}\colon  G \times M \to \mathds{R}^+$ as its innovation term. The cost function $\phi_{\mathring y}$ must be right invariant, and it should satisfy condition \ref{con_obsstability_phi_PD} of Theorem \ref{theo_observer_stability} (or condition \ref{con_observer_exp_phi} of Theorem \ref{theo_observer_exp_stability}) in order to guarantee asymptotic (or exponential) stability of the observer error. Designing such a cost function can be challenging since $M$ is an orbit in the product of different output spaces which can generally be a complicated manifold. In this section, based on the idea presented in \cite{mahony2013observers}, we propose a method for constructing a suitable cost function $\phi_{\mathring y}$ by employing single variable cost functions on the homogeneous output spaces $M_i$. Finding a suitable cost function on each output space is usually easy, especially when the output spaces are embedded in Euclidean spaces.

\begin{proposition} \label{lemma_cost}
 \cite{mahony2013observers} Suppose $f^i_{\mathring y_i}\colon M_i \to \mathds{R}^+,~y_i \mapsto f^i_{\mathring y_i}(y_i)$ are single variable $C^2$ cost functions on $M_i,~i=1,\ldots,n$. Corresponding to each $f^i_{\mathring y_i}$, construct a cost function $\phi^i_{\mathring y_i}\colon G \times M_i \to \mathds{R}^+$ using $\phi^i_{\mathring y_i}(\hat X, y_i):=f^i_{\mathring y_i}(h_i(\hat X^{-1}, y_i))$. Obtain the cost function $\phi_{\mathring y} (\hat X,y):=\sum_{i=1}^n {\phi^i_{\mathring y_i}(\hat X,y_i)}$.
\begin{enumerate}
\labitem{(a)}{con_prop_cost_phi_invar}{The cost function $\phi_{\mathring y}$ is right invariant in the sense defined in part \ref{con_obsstability_phi_invar} of Theorem \ref{theo_observer_stability}.}
\labitem{(b)}{con_prop_cost_phi_PD}{Assume that each $f^i_{\mathring y_i},~i=1,\ldots,n$ is locally positive definite around $\mathring y_i \in M_i$.  Assume moreover that $\bigcap_{i=1}^n {\text{stab}_{h_i}(\mathring{y}_i)}=\{ I \}$ where $\text{stab}_{h_i} (\mathring{y}_i)$ denotes the stabilizer of $\mathring{y}_i$ with respect to the action $h_i$, defined by $\text{stab}_{h_i}(\mathring{y}_i):= \{X \in G :~ h_i(X,\mathring{y}_i)=\mathring{y}_i \}$. Then $\phi_{\mathring y}(.,\mathring y):G \to \mathds{R}^+$ is locally positive definite around $I\in G$.}
\labitem{(c)}{con_prop_cost_phi_diff_hess}{If $\text{D} f^i_{\mathring y_i}(\mathring{y}_i)=0$ for all $i=1,\ldots,n$ then $\text{D}_1\phi_{\mathring y}(I,\mathring y)=0$. If additionally the Hessian operators $\text{Hess} f^i_{\mathring y_i}(\mathring{y}_i)\colon  T_{\mathring y_i} M_i \to T^*_{\mathring y_i} M_i$ are symmetric positive definite for all $i=1,\ldots,n$ and $\bigcap_{i=1}^n {T_I \text{stab}_{h_i}(\mathring{y}_i)}=\{ 0 \}$, then $\text{Hess}_1 \phi_{\mathring y}(I,\mathring y)$ is symmetric positive definite.}\carrew
%\labitem{(d)}{con_prop_cost_phi_compact}{If the cost functions $f^i_{\mathring y_i},~i=1,\ldots,n$ have compact sublevel sets and $\bigcap_{i=1}^n \text{stab}_{h_i}(\mathring{y}_i)$ is compact, then $\phi_{\mathring y}(.,\mathring y)\colon G\to \mathds{R}^+$ has compact sublevel sets.}
\end{enumerate}
\end{proposition}

Proof of Proposition \ref{lemma_cost} is given in the Appendix. Proposition \ref{lemma_cost} suggests a systematic method to construct a cost function which satisfies the requirements of Theorem \ref{theo_observer_stability} or Theorem \ref{theo_observer_exp_stability}. The differential of this function can be employed to design the innovation term of the observer. We will illustrate this method with an example in section \ref{sec_SE3_exmple}.

The method proposed by Proposition \ref{lemma_cost} to construct the cost function $\phi_{\mathring y}$ is basically different from the one presented in \cite[Proposition 2]{khosravian2013bias}. The method proposed in \cite{khosravian2013bias} employs invariant cost functions on $M_i \times M_i$ while the method presented here only requires single variable cost functions on each $M_i$. Implementability of the proposed observer in \cite{khosravian2013bias} is guaranteed when the homogeneous output spaces are reductive. The method presented in this paper guarantees the implementability of resulting observer without imposing any reductivity condition.

The condition $\bigcap_{i=1}^n {\text{stab}_{h_i}(\mathring{y}_i)}=\{ I \}$ (imposed in part \ref{con_prop_cost_phi_PD} of Proposition \ref{lemma_cost}) is sufficient to ensure $\bigcap_{i=1}^n {T_I \text{stab}_{h_i}(\mathring{y}_i)}=\{ 0 \}$ (imposed in part \ref{con_prop_cost_phi_diff_hess} of the Proposition). This condition can be interpreted as an observability criterion. In particular, for the attitude estimation problem with vectorial measurements, this condition is equivalent to the availability of two or more non-collinear reference vectors \cite{khosravian2013bias}. As will be discussed in the next section, for the pose estimation problem with landmark measurements, this condition corresponds to the availability of three or more landmarks which are not located on the same line.

The method presented in this paper suits the systems with constant reference outputs. Time varying reference outputs have been investigated in \cite{trumpf2012analysis,grip2012attitude,batista2012ges,Khosravian10} for attitude estimation problem on $\textsf{SO}(3)$. Nevertheless, in most practical cases, the reference outputs are approximately constant \cite{Mahony08,vasconcelos2010nonlinear,bras2011nonlinear} and the proposed observer design methodology applies.

\vspace{-1mm}
\section{Example: Pose Estimation Using Biased Velocity Measurements} \label{sec_SE3_exmple}
Recalling the pose estimation problem discussed in Example \ref{example_SE3}, here we employ our observer (\ref{observer_left_dyn})-(\ref{b_hat_dy}) to derive the pose estimators proposed in \cite{hua2011observer} and \cite{vasconcelos2010nonlinear} and we generalize them.

Apart from the semi-direct product representation of $\textsf{SE}(3)$ discussed in Example \ref{example_SE3}, it is known that $\textsf{SE}(3)$ has also a matrix Lie group representation as a subgroup of $\textsf{GL}(4)$ (see e.g. \cite{hua2011observer}). We use this matrix Lie group representation only to interpret the required boundedness conditions (see Assumption \ref{assumption_bounded}) but we employ the semi-direct product representation to derive the observer formulas (see remark \ref{rem_matrix_vs_general}). The Lie group homomorphism $\Phi$ which maps an element $(R,p) \in \textsf{SE}(3)$ to its corresponding matrix representation in $\textsf{GL}(4)$ is given by;
$
\Phi: (R,p) \mapsto \left[
  \begin{array}{cc}
    R & p \\
    0 & 1 \\
  \end{array}
\right].
$
The Frobenius norm of $\Phi((R,p)) \in \textsf{GL}(4)$ is given by $\|\Phi((R,p))\|^2=\text{tr}(\Phi((R,p))^\top \Phi((R,p)))=4+\|p\|^2$. Hence, $\Phi((R(t),p(t)))$ is bounded if $p(t)$ is bounded. Similarly, one can verify that $\Phi((R(t),p(t)))^{-1}$ and $\text{cond}((R(t),p(t)))$ are bounded (uniformly in $t_0$) if $p(t)$ is bounded (uniformly in $t_0$). This characterizes the boundedness conditions imposed by Assumption \ref{assumption_bounded} and part \ref{con_observer_exp_X_cond} of Theorem \ref{theo_observer_exp_stability}.

From here after, we only consider the semi-direct product representation of $\text{SE}(3) \simeq \textsf{SO}(3) \ltimes \mathds{R}^3$. We aim to employ the observer developed in section \ref{sec_observer_design} and use the guidelines presented in section \ref{sec_realizability_grad} to design an observer to estimate the pose $X=(R,p)$ and the bias $b=(b_\omega,b_v)$. Let us first evaluate the observability condition imposed by part \ref{con_prop_cost_phi_PD} and \ref{con_prop_cost_phi_diff_hess} of Proposition \ref{lemma_cost}. We have $\bigcap_{i=1}^n \text{stab}_{h_i}(\mathring{y}_i)=\{(R,p)\in \textsf{SE}(3):~R^\top \mathring{y}_i-R^\top p=\mathring{y}_i,~i=1,\ldots,n \}=\{(R,p)\in \textsf{SE}(3):~R^\top p=R^\top \mathring{y}_i-\mathring{y}_i,~R( \mathring{y}_i-\mathring{y}_j)=\mathring{y}_i-\mathring{y}_j ~i,j=1,\ldots,n,~i \ne j\}$ which implies that $\mathring y_i-\mathring y_j$ is an eigenvector of $R$.
Hence, a necessary and sufficient condition which guarantees $\bigcap_{i=1}^n \text{stab}_{h_i}(\mathring{y}_i)=\{ (I_{3 \times 3},0_3) \}$ (and consequently $\bigcap_{i=1}^n T_{(I,0)} \text{stab}_{h_i}(\mathring{y}_i)=\{ (0_{3\times 3},0_3) \}$) is the existence of at least three reference outputs $\mathring y_i,\mathring y_j,\mathring y_k$ such that $\mathring y_i-\mathring y_j$ is not parallel to $\mathring y_j-\mathring y_k$. Note that this condition is independent of the choice of inertial frame. Specifically, when landmark measurements are employed to provide outputs $y_i,~i=1,\ldots,n$, this condition is equivalent to the existence of at least three landmarks which are not located on the same line \cite{vasconcelos2010nonlinear,hua2011observer}.

In order to design the innovation terms of the estimator (\ref{observer_left_dyn})-(\ref{b_hat_dy}), we resort to choose a basis for each tangent space to obtain matrix representations for the linear mappings $K_{\mathring y},\Gamma,T_I L_{\hat X}^*$ and use simple matrix calculus. For the sake of clarity, we denote the matrix representation of a linear mapping $F\colon U \to W$ with respect to the basis $\{ \text{u} \}$ for its domain and basis $\{ \text{w} \}$ for its co-domain by the notation $[\![ F ]\!]_\text{u}^\text{w}$. Also, the $\mathds{R}^n$ representation of a vector $a \in U$ with respect to the basis $\{ \text{u} \}$ is denoted by $[\![ a ]\!]_\text{u}$. Denote the standard bases of $\mathds{R}^3$ and $\mathfrak{so}(3)$ by $\{ \text{e}\}$ and $\{\text{e}_\times\}$, respectively. Using these bases, one can obtain a standard basis for $\mathfrak{se}(3)$ denoted by $\{\overline{\text{e}}\}$. We obtain a basis for $T_{(\hat R, \hat p)} \textsf{SE}(3)$ using the right translation of $\{\overline{\text{e}}\}$. Denote this basis of $T_{(\hat R, \hat p)} \textsf{SE}(3)$ by
$\{ \overline{\text{e}} \hat X \}$ and its corresponding dual basis of $T^*_{(\hat R, \hat p)} \textsf{SE}(3)$ by $\{ ( \overline{\text{e}} \hat X)^* \}$.

In order to use Proposition \ref{lemma_cost}, we start by designing suitable costs $f^i_{\mathring y_i}:M_i \to \mathds{R}^+$. A simple cost function is constructed by $f^i_{\mathring y_i} (y_i):=\frac{k_i}{2} \| y_i - \mathring y_i \|^2,~k_i>0$ where $\|.\|$ denotes the Euclidean distance. It is straight forward to verify that $f^i_{\mathring y_i}$ satisfies the requirements imposed by part \ref{con_prop_cost_phi_diff_hess} of Proposition \ref{lemma_cost}, i.e. $\text{D} f^i_{\mathring y_i}(\mathring y_i)=0$ and $\text{Hess} f^i_{\mathring y_i}(\mathring y_i)$ is symmetric positive definite. The cost functions $\phi^i_{\mathring y_i}\colon \textsf{SE}(3) \times M_i \to \mathds{R}^+,~i=1,\ldots,n$ are obtained as $\phi^i_{\mathring y_i}(\hat X,y_i)=\frac{k_i}{2} \| h_i((\hat R,\hat p)^{-1},y_i) - \mathring y_i \|^2=\frac{k_i}{2} \| \hat R y_i + \hat p - \mathring y_i \|^2$. Denoting an arbitrary element of $\mathfrak{se}(3)$ by $(\hat \Omega, \hat V)$, we have $T_{(I,0)} R_{(\hat R,\hat p)} [(\hat \Omega,\hat V)]=(\hat \Omega \hat R, \hat \Omega \hat p+V)\in T_{(\hat R,\hat p)} \textsf{SE}(3)$. One can obtain $\text{D}_1 \phi_{\mathring y}((\hat R,\hat p),y)\colon T_{(\hat R,\hat p)} \textsf{SE}(3) \to \mathds{R}$ as
\begin{align} \label{example_pose_Dphi}
\text{D}_1 \phi_{\mathring y}((\hat R,\!\hat p),\!y)[(\hat \Omega \hat R, \!\hat \Omega \hat p\!+\!\hat V)]\!=\!\sum\nolimits_{i=1}^n{\!k_i \alpha_i^\top (\hat \Omega \hat R  y_i\!+\!\hat \Omega \hat p\!+\!\!\hat V)}.
\end{align}
where $\alpha_i:=(\hat R y_i+\hat p-\mathring y_i) \in \mathds{R}^3$. The $\mathds{R}^6$ representation of $\text{D}_1 \phi_{\mathring y}(\hat X,y) \in T^*_{\hat X} \textsf{SE}(3)$ is the transpose of the matrix representation of $\text{D}_1 \phi_{\mathring y}(\hat X,y): T_{\hat X} \textsf{SE}(3) \to \mathds{R}$, i.e. $[\![ \text{D}_1 \phi_{\mathring y}(\hat X,y) ]\!]_{( \overline{\text{e}} \hat X)^*}=\left ( \![\![ \text{D}_1 \phi_{\mathring y}(\hat X,y) ]\!]_{\overline{\text{e}} \hat X}^1\right )^\top$. Employing (\ref{example_pose_Dphi}) and using the simplifications given in the Appendix, we obtain
\begin{align}
\label{D_phi_original} &[\![ \text{D}_1 \phi_{\mathring y}(\hat X,y) ]\!]_{ \overline{\text{e}} \hat X}^1=\sum\nolimits_{i=1}^n{k_i  [ \mathring y_i^\top (\hat R y+\hat p)_\times ,\alpha_i^\top]}
\end{align}
We choose $[\![ K_{\mathring y}(\hat X,y,\hat b,u_y,t)]\!]_{( \overline{\text{e}} \hat X)^*}^{ \overline{\text{e}} \hat X}=  \text{diag}(k_\omega I_{3 \times 3},k_v I_{3 \times 3})$ where $k_\omega,k_v$ are positive scalars and ensure that the resulting gain mapping $K_{\mathring y}(\hat X, y,\hat b,u_y,t)\colon T^*_{(\hat R,\hat p)} \textsf{SE}(3) \to T_{(\hat R,\hat p)} \textsf{SE}(3)$ is uniformly positive definite. Using (\ref{D_phi_original}) we have
\begin{align}
\nonumber [\![ K_{\mathring y}(.)[\text{D}_1 \phi_{\mathring y}&(\hat X,y)] ]\!]_{ \overline{\text{e}}\hat X}=  [\![ K_{\mathring y}(.)]\!]_{( \overline{\text{e}} \hat X)^*}^{ \overline{\text{e}} \hat X} [\![ \text{D}_1 \phi_{\mathring y}(\hat X,y) ]\!]_{( \overline{\text{e}} \hat X)^*}\\
\label{se3_innovation_expressed} &=\left (\sum\nolimits_{i=1}^n{k_i [k_\omega \mathring y_i^\top (\hat R y_i+\hat p)_\times ,k_v \alpha_i^\top]}\right )^\top
\end{align}
where the argument $(\hat X,y,\hat b,u_y,t)$ of $K_{\mathring y}$ has been omitted for brevity. We use (\ref{R_n_to_se3X}) of Lemma \ref{lem:se3:convesion} given in the Appendix to obtain
\begin{align} \label{se3_innovation}
&K_{\mathring y}(.)[\text{D}_1 \phi_{\mathring y}(\hat X,y)]\\
\nonumber &=\!\sum\nolimits_{i=1}^n{\!k_i \left ( -k_\omega ((\hat R y_i+\hat p)_\times \mathring y_i)_\times \hat R,  -k_\omega ((\hat R y_i+\hat p)_\times \mathring y_i)_\times \hat p\!+\! k_v \alpha_i \right)}
\end{align}
Choosing  the gain $[\![ \Gamma ]\!]_{\overline{\text{e}}^*}^{\overline{\text{e}}}=\text{diag}(\gamma_\omega I_{3 \times 3},\gamma_v I_{3 \times 3})$, we have
\begin{align}
\nonumber &[\![ \Gamma \!\!\circ\! T_{\!I} L_{\!\hat X}^* [\text{D}_1 \phi_{\mathring y} (\hat X\!,\! y)] ]\!]_{\overline{\text{e}}}=[\![ \Gamma ]\!]_{\overline{\text{e}}^*}^{\overline{\text{e}}} \left ( \![\![ T_I L_{(\hat R, \hat p)} ]\!]_{\overline{\text{e}}}^{ \overline{\text{e}} \hat X} \right )^\top [\![ \text{D}_1 \phi_{\mathring y}(\hat X,y) ]\!]_{( \overline{\text{e}} \hat X)^*}\\
\label{example:tmpx} &=\left[
  \begin{array}{cc}
    {\!\!\!\!\gamma_\omega I_{3 \times 3}} & {\!\!\!0_{3\times 3}} \\
    {\!\!0_{3\times 3}} & {\!\!\!\!\gamma_v I_{3 \times 3}} \\
  \end{array}
\!\!\!\!\right]
\!\left[
  \begin{array}{cc}
    {\!\!\!\!\!\!\!\hat R} & {\!\!\!\!\!\! 0_{3\times 3} } \\
    {\!\!\! \hat p_\times \hat R} & {\hat R} \\
  \end{array}
\!\!\!\!\right]^\top
\!\left[
  \begin{array}{c}
    {\!\!\!-\sum_{i=1}^n{k_i (\hat R y_i+\hat p)_\times \mathring y_i}}\\
    {\sum_{i=1}^n{k_i\alpha_i}}\\
  \end{array}
\!\!\!\right]\\
\nonumber &=\sum\nolimits_{i=1}^n{k_i\left[
  \begin{array}{c}
    {\gamma_\omega {y_i}_\times (\hat R^\top \mathring y_i -\hat R^\top \hat p)}\\
    {\gamma_v \hat R^\top \alpha_i}\\
  \end{array}
\right]}.
\end{align}
where the term $[\![ T_I L_{(\hat R, \hat p)} ]\!]_{\overline{\text{e}}}^{ \overline{\text{e}} \hat X}$ has been computed in the Appendix. One can employ (\ref{R_n_to_se3}) of Lemma \ref{lem:se3:convesion} given in the Appendix to obtain
\begin{align}
\label{se3_bia_innovation} \!\!\!\!\Gamma \!\!\circ\! T_{\!I} L_{\!\hat X}^* [\text{D}_1 \phi_{\mathring y} (\hat X,\! y)]\!=\!\!\!\sum\nolimits_{i=1}^n{ \!\!k_i ( \gamma_\omega ({y_i}_\times (\hat R^\top \mathring y_i \!-\!\hat R^\top \!\hat p))_\times , \gamma_v \hat R^\top \!\alpha_i)}
\end{align}
Using (\ref{se3_innovation}) and (\ref{se3_bia_innovation}), the observer is summarized as
\begin{subequations} \label{example_SE3_observer_MD}
\begin{align}
\label{example_SE3_observer_MD_R}  &\!\!\dot{\hat R}=\hat R (\Omega_y-\hat b_\omega)+k_\omega \sum\nolimits_{i=1}^n{ k_i((\hat R y+\hat p)_\times \mathring y_i)_\times} \hat R \\
\label{example_SE3_observer_MD_p} &\!\!\dot{\hat p}\!=\!\hat R(V_y\!-\!\hat b_v)\!+\!\!\sum_{i=1}^n{ \!k_i (k_\omega ((\hat R y+\hat p)_\times \mathring y_i)_\times \hat p\!-\!k_v (\hat R y_i \!+\!\hat p\!-\!\mathring y_i))}\\
&\!\! \dot{\hat b}_w=\gamma_\omega \sum\nolimits_{i=1}^n{ k_i({y_i}_\times (\hat R^\top \mathring y_i -\hat R^\top \hat p))_\times}\\
\label{example_SE3_observer_MD_bv} &\!\!\dot{\hat b}_v= \gamma_v (\hat R^\top \hat p -\hat R^\top \mathring y_i+y_i)
\end{align}
\end{subequations}
Notice that the resulting observer formulas (\ref{example_SE3_observer_MD_R})-(\ref{example_SE3_observer_MD_bv}) do not depend on the chosen basis. Omitting the bias estimator, the group estimator (\ref{example_SE3_observer_MD_R})-(\ref{example_SE3_observer_MD_p}) has a similar form as the gradient-like observer proposed in \cite[equation (35)]{hua2011observer} since the chosen gain mapping $K_{\mathring y}$ is symmetric positive definite and yields a gradient innovation term.

The pose estimator of \cite{vasconcelos2010nonlinear} has a different form from (\ref{example_SE3_observer_MD}). Here, we derive the observer of \cite{vasconcelos2010nonlinear} by choosing different gain mappings and output maps. Similar to \cite[equation (8)]{vasconcelos2010nonlinear}, consider the new set of outputs $z_j,~j=1,\ldots,n$ given by
\begin{subequations} \label{z_SE3}
\begin{align}
z_j &:=\sum\nolimits_{i=1}^{n-1}{a_{ij}(y_{i+1}-y_i)},~~~~~~j=1,\ldots,n-1\\
z_n &:=-\frac{1}{n} \sum\nolimits_{i=1}^n{y_i}.
\end{align}
\end{subequations}
We assume that $a_{ij} \in \mathds{R}$ are such that the matrix $A:=[a_{ij}]\in \mathds{R}^{(n-1)\times (n-1)}$ is full rank. This requirement guarantees that no information is lost by applying the linear transformation (\ref{z_SE3}) to the measurements. Substituting $y_i$ from (\ref{SE3_measurement}) into (\ref{z_SE3}) and defining new reference outputs $\mathring z_j:=\sum_{i=1}^{n-1}{a_{ij}(\mathring y_{i+1}-\mathring y_i)},~j=1,\ldots,n-1$, $\mathring z_n=-\frac{1}{n} \sum_{i=1}^n{\mathring y_i}$ yields
\begin{subequations}
\begin{align}
\label{z_SE3_j} z_j &=g_j((R,p),\mathring z_j):=R^\top \mathring z_j,~~~~~~~~j=1,\ldots,n-1\\
z_n &=g_n((R,p),\mathring z_n):=R^\top \mathring z_n + R^\top p
\end{align}
\end{subequations}
where $g_j,~j=1,\ldots,n$ are right output actions of $G$. Consider the new combined output $z:=(z_1,\ldots,z_n)$ and the combined reference output $\mathring z:=(\mathring z_1,\ldots,\mathring z_n)$. One can show that the necessary and sufficient condition for $\bigcap_{j=1}^n \text{stab}_{g_j}(\mathring{z}_j)=\{ I \}$ is the existence of at least two non-collinear reference outputs $\mathring z_j,\mathring z_k$. Assuming that $A=[a_{ij}]$ is invertible, it is straight forward to show that the above mentioned condition on $\mathring z$ is equivalent to the condition on $\mathring y$ we derived before.

We employ the cost functions $f^j_{\mathring z_j} (z_j):=\frac{k_j}{2} \| z_j - \mathring z_j \|^2,~k_j>0$ and
we choose the gain mappings
$[\![ K_{\mathring z}(\hat X,y,\hat b,u_y,t)]\!]_{(\overline{\text{e}} \hat X)^*}^{\overline{\text{e}} \hat X}=  \text{diag}(k_w I_{3\times 3},k_v I_{3\times 3})+\text{diag}(0_{3\times 3},(\hat R(\Omega_y-\hat b_\omega))_\times)$ and $[\![ \Gamma ]\!]_{\overline{\text{e}}^*}^{\overline{\text{e}}}=\text{diag}(\gamma_\omega I_{3 \times 3},\gamma_v I_{3 \times 3})$.
It is easy to verify that this choice of cost functions and gain mappings satisfies the requirements of our
method. Notice that $K_{\mathring z}$ is non-symmetric and depends also on $\Omega_y$ and
$\hat b_\omega$ unlike the previous part.
In particular, this implies that the observer innovation is not a gradient innovation.
Nevertheless, the symmetric part of $K_{\mathring z}$ is $ \text{diag}(k_w I_{3\times 3},k_v I_{3\times 3})$ which implies that the resulting gain mapping $K_{\mathring z}$ is uniformly positive definite. Following the same procedure as was done to derive (\ref{example_SE3_observer_MD}), we obtain the following observer.
\begin{subequations} \label{example_SE3_observer_Carlos}
\begin{align}
\dot{\hat R}&\!=\!\hat R (\Omega_y\!-\!\hat b_\omega\!)\!-\! k_\omega k_n (\hat p_\times \!\mathring z_n\!)_\times \hat R \!+\!k_\omega \sum\nolimits_{j=1}^{n}{\!k_j  \hat R \big(\!(\!\hat R\!^\top \!\mathring z_j\!)_\times z_j\!\big)_{\!\times}} \\
\nonumber \dot{\hat p}&\!=\!\hat R(V_y\!-\!\hat b_v\!)\! +\!k_n \big(k_v I_{3\times 3}+ (\hat R(\Omega_y\!-\!\hat b_\omega))_\times \big) (\hat R z_n-\hat p-\mathring z_n) \\
\label{example_SE3_observer_Carlos_p_hat} &~~~~ - k_\omega k_n (\hat p_\times \mathring z_n)_\times \hat p+k_\omega \sum\nolimits_{j=1}^{n}{k_j \big( (\hat R  {z_j})_\times} \mathring z_j\big)_{\times} \hat p\\
\dot{\hat b}_w&\!=\! \gamma_\omega k_n \Big( \hat R\!^\top \hat p_\times (\hat R z_n\!-\!\hat p) \Big)_\times \!+\gamma_\omega \sum\nolimits_{j=1}^{n}{k_j  \Big( (\hat R^\top \mathring z_j)_\times z_j\Big)_\times }\\
\dot{\hat b}_v&= -\gamma_v k_n \hat R^\top (\hat R z_n- \hat p-\mathring z_n)
\end{align}
\end{subequations}
In \cite{vasconcelos2010nonlinear}, it is assumed that the origin of inertial frame is located at the geometric center of the landmarks. In this case we have $\mathring z_n=0$ which simplifies the observer (\ref{example_SE3_observer_Carlos}) to the observer designed in \cite{vasconcelos2010nonlinear}\footnote{Here, the position vector $p$ is expressed in the inertial frame but in \cite{vasconcelos2010nonlinear} the position vector is expressed in the body-fixed frame. One can transform the system of \cite{vasconcelos2010nonlinear} to the form presented here using the change of variable $p \mapsto Rp$.}. Compared to \cite{vasconcelos2010nonlinear}, the observer (\ref{example_SE3_observer_Carlos}) has the advantage that it is well-defined even if only some of the measurements $y_i$ are unavailable at some period of time. In this case, the reference output $\mathring z_n$ can be recalculated using the reference outputs corresponding to the remaining available measurements. Also, we only require $A=[a_{ij}]$ to be full rank but \cite{vasconcelos2010nonlinear} necessarily requires that $a_{ij}$ are chosen such that $[\mathring z_1,\ldots,\mathring z_{n-1}][\mathring z_1,\ldots,\mathring z_{n-1}]^\top=I_{3\times 3}$. For practical implementation purpose, discrete time representation of the observers could be obtained using standard structure preserving numerical integration methods \cite{iserles2000lie}.

\section{Conclusion} \label{sec_conclusion}
We investigate the problem of observer design for invariant systems on finite-dimensional real connected Lie groups where the measurement of system input is corrupted by an unknown constant bias.
We show that the corresponding standard error dynamics are non autonomous in general.
We propose an observer design methodology that guarantees the uniform local asymptotic (or exponential)
convergence of the observer trajectories to the system trajectories.
We employ a gain mapping acting on the differential of a cost function to design the innovation term of the
group estimator. The bias estimator is then designed using a Lyapunov method.
The notion of homogeneous output spaces is generalized to multiple outputs, each of which is modeled
via a right action of the Lie group on an output space. A systematic method for constructing invariant
cost functions on Lie groups is proposed, yielding implementable innovation terms for the observer.
A verifiable condition on the stabilizer of the reference outputs associated with the output spaces
ensures the stability of the observer. This condition is consistent with the observability criterion
discussed in \cite{Lageman09}.
Our proposed method omits the limiting reductivity condition imposed in the
authors' previous work \cite{khosravian2013bias,lageman2008observer}.
As a case study, pose estimation on the Lie group $\textsf{SE}(3)$ was investigated where our observer design methodology unifies the state-of-the-art
pose estimators of \cite{vasconcelos2010nonlinear} and \cite{hua2011observer} into a single framework that applies to any invariant kinematic system on a Lie-group. Extension of the proposed observer design methodology to the (co)tangent bundle of a Lie group could be considered by assigning a Lie group structure to the (co)tangent bundle noting that the (co)tangent bundle is trivial (see e.g. \cite{saccon2013second}).

\section*{APPENDIX}
\subsection*{Proof of Lemma \ref{lemma_Dphi_simplify}:}
The right-invariance property of $\phi_{\mathring y}$ implies $\phi_{\mathring y}(\hat X,y)=\phi_{\mathring y} \circ R_{X^{-1}} (\hat X,y)$. Differentiating both sides in an arbitrary direction $v \in T_{\hat X} G$ and using the chain rule we obtain $\text{D}_1 \phi_{\mathring y}(\hat X,y) [v]=\text{D}_1 \phi_{\mathring y}(E,\mathring y) \circ T_{\hat X} R_{X^{-1}} [v]$.
Since $v$ is arbitrary and by using the duality we have $\label{proof_Dphi_temp} \text{D}_1 \phi_{\mathring y}(\hat X,y)=T_{\hat X} R_{X^{-1}}^* [\text{D}_1 \phi_{\mathring y} (E,\mathring y)]$ which proves (\ref{D_phi_E2X}). Applying $(T_{\hat X} R_{X^{-1}})^{-1}=T_{E} R_X$ from the right to both sides of $\text{D}_1 \phi_{\mathring y}(\hat X,y)=\text{D}_1 \phi_{\mathring y}(E,\mathring y) \circ T_{\hat X} R_{X^{-1}}$ yields (\ref{D_phi_X2E}). \carre

\subsection*{Proof of Proposition \ref{lemma_cost}:}
\subsubsection*{Part \ref{con_prop_cost_phi_invar}} For any arbitrary $Z\in G$ we have $\phi_{\mathring y}(\hat XZ,h(Z,y))=\sum_{i=1}^n {f^i_{\mathring y_i}\left(h_i((\hat X Z)^{-1},h_i(Z , y))\right)}=\sum_{i=1}^n {f^i_{\mathring y_i}(h_i(Z Z^{-1} \hat X^{-1},y))}=\sum_{i=1}^n {f^i_{\mathring y_i}(h_i(\hat X^{-1},y))}=\phi_{\mathring y}(\hat X,y)$ which shows that $\phi_{\mathring y}$ is right invariant.

\subsubsection*{Part \ref{con_prop_cost_phi_PD}}
Since $f^i_{\mathring y_i}(y_i)$ is positive definite around $y_i=\mathring y_i$, there exists a neighborhood $N_i \subset M_i$ of $\mathring y_i$ such that $f^i_{\mathring y}(y_i) \ge 0$ and $f^i_{\mathring y_i}(y_i)=0 \Rightarrow y_i=\mathring y_i$ for all $y_i \in N_i$. Corresponding to each $N_i$, define the set $\overline{N}_i:=\{E\in G: h_i(E^{-1},\mathring y_i) \in N_i\} \subset G$ and consider the set $\overline{N}:=\bigcap_{i=1}^n{\overline{N}_i}$. It is easy to verify that $\overline{N} \subset G$ is a neighborhood of $I$ and we have $\phi_{\mathring y}(E,\mathring y)=\sum_{i=1}^n {f^i_{\mathring y_i}(h_i(E^{-1},\mathring y_i))} \ge 0$ for all $E \in \overline{N}$. Moreover, for any $E \in \overline{N}$, $\phi_{\mathring y}(E,\mathring y)=\sum_{i=1}^n {f^i_{\mathring y_i}(h_i(E^{-1},\mathring y_i))}=0$ yields $f^i_{\mathring y_i}(h_i(E^{-1},\mathring y_i))=0$ for all $i=1,\ldots,n$. This in turn implies that $h_i(E^{-1},\mathring y_i)=\mathring y_i,~i=1,\ldots,n$ and hence $E \in \bigcap_{i=1}^n\text{stab}_{h_i}(\mathring y_i)$. We assumed $\bigcap_{i=1}^n \text{stab}_{h_i}(\mathring y_i)=\{ I \}$ which ensures that $E=I$ and hence $\phi_{\mathring y}(E,\mathring y)$ is positive definite on $\overline{N}$.

\subsubsection*{Part \ref{con_prop_cost_phi_diff_hess}}
Define the map $h_{\mathring{y}_i}\colon  G\to M_i$ by $h_{\mathring{y}_i} X :=h_i(X,\mathring{y}_i)$. Differentiating both sides of $\phi_{\mathring y}(E,\mathring y)=\sum_{i=1}^n {f^i_{\mathring y_i}(h_i(E^{-1},\mathring y_i)}$ in an arbitrary direction $v\in T_E G$ and using the chain rule we have $\text{D}_1 \phi_{\mathring y}(E,\mathring y)[v]=- \sum_{i=1}^n {\text{D} f^i_{\mathring y_i}(h(E^{-1},\mathring{y}_i)) \circ T_{E^{-1}} h_{\mathring y_i} \circ T_I L_{E^{-1}} \circ T_E R_{E^{-1}}[v]}$. Evaluating the later relation at $E=I$ and omitting the arbitrary argument $v$ we obtain $\text{D}_1 \phi_{\mathring y}(I,\mathring y)=- \sum_{i=1}^n {\text{D} f^i_{\mathring y_i}(\mathring{y}_i)\circ T_I h_{\mathring{y}_i}}$. Hence, $\text{D} f^i_{\mathring y_i}(\mathring{y}_i)=0,~i=1,\ldots,n$ implies $\text{D}_1 \phi_{\mathring y}(I,\mathring y)=0$.
Under this condition, standard computations shows that $\text{Hess}_1 \phi_{\mathring y}(I,\mathring y)=\sum_{i=1}^n {\text{Hess}_1 \phi_i(I,\mathring y_i)}=\sum_{i=1}^n {T_I h_{\mathring{y}_i}^* \circ \text{Hess} f^i_{\mathring y_i}(\mathring{y}_i) \circ T_I h_{\mathring{y}_i}}$ where $T_I h_{\mathring{y}_i}^*\colon  T^*_{\mathring y_i} M_i \to T^*_I G$ denotes the dual of $T_I h_{\mathring{y}_i}$. If all of $\text{Hess} f^i_{\mathring y_i}(\mathring{y}_i),~i=1,\ldots,n$ are symmetric positive definite, then $\text{Hess}_1 \phi_{\mathring y}(I,\mathring y)$ is symmetric positive semi definite with $\text{ker}(\text{Hess}_1 \phi_{\mathring y}(I,\mathring y))=\bigcap_{i=1}^n {\text{ker}(T_I h_{\mathring{y}_i})}$. Since, $\text{ker}(T_I h_{\mathring{y}_i})=T_I \text{stab}_{h_i}(\mathring{y}_i)$, we have $\bigcap_{i=1}^n {\text{ker}(T_I h_{\mathring{y}_i})}=\bigcap_{i=1}^n {T_I \text{stab}_{h_i}(\mathring{y}_i)}$ which is assumed to be $\{ 0 \}$. Consequently, $\text{ker}(\text{Hess}_1 \phi_{\mathring y}(I,\mathring y))=\{ 0 \}$ which implies that $\text{Hess}_1 \phi_{\mathring y}(I,\mathring y)$ is full rank and hence symmetric positive definite.
\carre

\subsection*{Computing $[\![ \text{D}_1 \phi_{\mathring y}(\hat X,y) ]\!]_{ \overline{\text{e}} \hat X}^1$ employed in (\ref{D_phi_original}): }
\vspace{-0.1cm}
The standard basis for $\mathds{R}^3$ is given by $\{ \text{e}\}:=\{\text{e}^1,\text{e}^2,\text{e}^3\}$. The standard basis for $\mathfrak{so}(3)$ is obtained as $\{\text{e}_\times\}:=\{\text{e}^1_\times,\text{e}^2_\times,\text{e}^3_\times\}$ and a basis for $\mathfrak{se}(3)$ is represented by $\{\overline{\text{e}}\}:=\{(\text{e}^1_\times,0_3),(\text{e}^2_\times,0_3),(\text{e}^3_\times,0_3),(0_{3\times 3},\text{e}^1),(0_{3\times 3},\text{e}^2),(0_{3\times 3},\text{e}^3)\}$. A basis for $T_{(\hat R, \hat p)} \textsf{SE}(3)$ is obtained by right-translating the basis of $\mathfrak{se}(3)$ as 
\begin{align*}
\{  \overline{\text{e}} \hat X \}:=\{&(\text{e}^1_\times \hat R ,\text{e}^1_\times \hat p),(\text{e}^2_\times \hat R ,\text{e}^2_\times \hat p),(\text{e}^3_\times \hat R ,\text{e}^3_\times \hat p),\\
&(0_{3\times 3},\text{e}^1),(0_{3\times 3},\text{e}^2),(0_{3\times 3},\text{e}^3)\}.
\end{align*}
We employ the above basis to obtain
\begin{align*}
&[\![ \text{D}_1 \phi_{\mathring y}(\hat X,y) ]\!]_{ \overline{\text{e}} \hat X}^1\\
&=\sum_{i=1}^n{k_i \alpha_i^\top [ \text{e}^1_\times \hat R y_i+ \text{e}^1_\times \hat p,\text{e}^2_\times \hat R y_i+\text{e}^2_\times \hat p, \text{e}^3_\times \hat R y_i+\text{e}^3_\times \hat p, \text{e}^1, \text{e}^2, \text{e}^3]}\\
&=\sum_{i=1}^n{k_i  [ -\alpha_i^\top (\hat R y_i+\hat p)_\times,\alpha_i^\top]}=\sum_{i=1}^n{k_i  [ \mathring y_i^\top (\hat R y+\hat p)_\times ,\alpha_i^\top]}.
\end{align*}
\carre

\begin{lemma} \label{lem:se3:convesion}
Suppose that the $\mathds{R}^6$ representation of $u \in \mathfrak{se}(3)$ and $w \in T_{(\hat R,\hat p)} \textsf{SE}(3)$ with respect to the basis $\{ \overline{\text{e}} \}$ and $\{ \overline{\text{e}} \hat X \}$ are respectively given by $[\![ u ]\!]_{\overline{\text{e}}}=[u_\omega^\top,u_v^\top]^\top$ and $[\![w]\!]_{\overline{\text{e}}\hat X}=[w_\omega^\top,w_v^\top]^\top$ where $u_\omega,u_v,w_\omega,w_v \in \mathds{R}^3$. Then $u$ and $w$ can be written as in terms of their $\mathds{R}^6$ representation as follows.
\begin{align}
\label{R_n_to_se3} u&=({u_\omega}_\times, u_v)\\
\label{R_n_to_se3X} w&=({w_\omega}_{\times} \hat R,{w_\omega}_{\times}\hat p+w_v).
\end{align} \carrew
\end{lemma}

\textit{Proof: }
\vspace{-0.1cm}
\begin{align*}
w=&w_\omega^\top \text{e}^1 (\text{e}^1_\times \hat R ,\text{e}^1_\times \hat p)+w_\omega^\top \text{e}^2 (\text{e}^2_\times \hat R ,\text{e}^2_\times \hat p)+w_\omega^\top \text{e}^3 (\text{e}^3_\times \hat R ,\text{e}^3_\times \hat p)\\
& +w_v^\top \text{e}^1 (0_{3\times 3},\text{e}^1) +w_v^\top \text{e}^2 (0_{3\times 3},\text{e}^2) + w_v^\top \text{e}^3 (0_{3\times 3},\text{e}^3)\\
\nonumber =&\Big((w_\omega^\top \text{e}^1  \text{e}^1_\times+w_\omega^\top \text{e}^2  \text{e}^2_\times+w_\omega^\top \text{e}^3  \text{e}^3_\times) \hat R,\\
\nonumber &~~\big(w_\omega^\top \text{e}^1  \text{e}^1_\times \!+\! w_\omega^\top \text{e}^2  \text{e}^2_\times \!+\! w_\omega^\top \text{e}^3  \text{e}^3_\times \big) \hat p \!+\! w_v^\top \text{e}^1 \text{e}^1 \!+\! w_v^\top \text{e}^2 \text{e}^2 \!+\! w_v^\top \text{e}^3 \text{e}^3 \Big)\\
=&({w_v}_\times \hat R, {w_v}_\times \hat p+w_v)
\end{align*}
where we used the standard equation $a=a^\top \text{e}^1 \text{e}^1 +a^\top \text{e}^2 \text{e}^2+a^\top \text{e}^3 \text{e}^3$ once for $a=w_\omega$ and once for $a=w_v$ to obtain the last line. This proves (\ref{R_n_to_se3X}). Choosing $(\hat R,\hat p)=(I_{3\times 3},0)$, it is easy to verify that (\ref{R_n_to_se3}) holds too.\carre

\subsection*{Computing $[\![T_I  L_{(\hat R,\hat p)}]\!]_{\text{e}}^{\text{e} \hat X}$ employed in (\ref{example:tmpx}):}
Suppose $[a^\top,b^\top]^\top \in \mathds{R}^6$ as the first column of $[\![T_I  L_{(\hat R,\hat p)}]\!]_{\overline{\text{e}}}^{\overline{\text{e}} \hat X}$ and denote by $a_i,b_i,~i=1,\ldots 3$ the elements of $a,b\in \mathds{R}^3$. We have,
$
T_I  L_{\hat X}[(\text{e}^1_\times,0)]=(\hat R \text{e}^1_\times,0)=\sum_{i=1}^3 {a_i (\text{e}^i_\times \hat R, \text{e}^i_\times \hat p)+b_i(0,\text{e}^i)}= \sum_{i=1}^3{(a_i \text{e}^i_\times \hat R, a_i \text{e}^i_\times \hat p+b_i\text{e}^i)}
$.
This implies that $\hat R \text{e}^1_\times=\sum_{i=1}^3{a_i \text{e}^i_\times \hat R}$ and $0=\sum_{i=1}^3{ a_i \text{e}^i_\times \hat p+b_i\text{e}^i}$ which together form 6 linear equations with 6 unknowns. Solving this set of equations yields $a=\hat R \text{e}^1$ and $b=\hat p_\times \hat R \text{e}^1$. Consequently, the first column of $[\![T_I  L_{(\hat R,\hat p)}]\!]_{\overline{\text{e}}}^{\overline{\text{e}} \hat X}$ is given by $[(\hat R \text{e}^1)^\top,(\hat p_\times \hat R \text{e}^1)^\top]^\top$. One can use the same procedure as was explained above to obtain the second and third column of $[\![T_I  L_{(\hat R,\hat p)}]\!]_{\overline{\text{e}}}^{\overline{\text{e}} \hat X}$ as $[(\hat R \text{e}^2)^\top,(\hat p_\times \hat R \text{e}^2)^\top]^\top$ and $[(\hat R \text{e}^3)^\top,(\hat p_\times \hat R \text{e}^3)^\top]^\top$ respectively. Suppose $[c^\top,d^\top]^\top$ as the forth column of $[\![T_I  L_{(\hat R,\hat p)}]\!]_{\overline{\text{e}}}^{\overline{\text{e}} \hat X}$. We have,
$
T_I  L_{\hat X}[(0,\text{e}^1)]=(0,\hat R \text{e}^1)= \sum_{i=1}^3{(c_i \text{e}^i_\times \hat R, c_i \text{e}^i_\times \hat p+d_i\text{e}^i)}
$.
This implies that $0=\sum_{i=1}^3{c_i \text{e}^i_\times \hat R}$ and $\hat R \text{e}^1= \sum_{i=1}^3{ c_i \text{e}^i_\times \hat p+d_i\text{e}^i}$ which again form 6 linear equations with 6 unknowns. Solving this set of equations yields $c=0$ and $d=\hat R \text{e}^1$. Hence the forth column of $[\![T_I  L_{(\hat R,\hat p)}]\!]_{\overline{\text{e}}}^{\overline{\text{e}} \hat X}$ is given by $[0,(\hat R \text{e}^1)^\top]^\top$. We can apply the same procedure to obtain the fifth and sixth column as well. Combining all of the columns together yields
\begin{align*}
[\![T_I  L_{(\hat R,\hat p)}]\!]_{\overline{\text{e}}}^{\overline{\text{e}} \hat X}&=
\left[
\begin{array}{cccccc}
   {\hat R \text{e}^1} & {\hat R \text{e}^2} & {\hat R \text{e}^3} & {0_3} & {0_3} & {0_3} \\
  {\hat p_\times \hat R\text{e}^1} & {\hat p_\times \hat R\text{e}^2} & {\hat p_\times \hat R\text{e}^3} & {\hat R \text{e}^1} & {\hat R \text{e}^2} & {\hat R \text{e}^3} 
\end{array}
\right]\\
&=
\left[
\begin{array}{cc}
   {\hat R} &{0_{3\times 3}} \\
  {\hat p_\times \hat R} & {\hat R} 
\end{array}
\right].
\end{align*}

\vspace{-7mm}

\carre

%\vspace{-0.5mm}
%\section{Acknowledgment}

%\ifCLASSOPTIONcaptionsoff
%  \newpage
%\fi

%\vspace{-0.5mm}
\bibliographystyle{elsarticle-num}
\bibliography{librarysample}

\end{document}